# Engineering coherent interactions in molecular nanomagnet dimers

Arzhang Ardavan,[1]* Alice M. Bowen,[1] Antonio Fernandez,[2] Alistair J. Fielding,[2] Danielle Kaminski,[1] Fabrizio Moro,[2] Christopher A. Muryn,[2] Matthew D. Wise,[3] Albert Ruggi,[3] Eric J.L. McInnes,[2]* Kay Severin,[3] Grigore A. Timco,[2] Christiane R. Timmel,[4] Floriana Tuna,[2] George F.S. Whitehead,[2] & Richard E. P. Winpenny[2]*

1. Centre for Advanced Electron Spin Resonance, The Clarendon Laboratory, Department of Physics, University of Oxford, Parks Road, Oxford OX1 3PU, U.K. Tel: +44 (0)1865 272366; E-mail: arzhang.ardavan@physics.ox.ac.uk

2. School of Chemistry & Photon Science Institute, The University of Manchester, Oxford Road, Manchester, M13 9PL, U.K.

3. Institut des Sciences et Ingenierie Chimiques, Ecole Polytechnique Federale de Lausanne (EPFL), 1015 Lausanne, Switzerland.

4. Centre for Advanced Electron Spin Resonance, Inorganic Chemistry Laboratory, South Parks Road, University of Oxford, OX1 3QR, U.K.

*Abstract*

**Proposals for systems embodying condensed matter spin qubits cover a very wide range of length scales, from atomic defects in semiconductors all the way to micron-sized lithographically-defined structures. Intermediate scale molecular components exhibit advantages of both limits: like atomic defects, large numbers of identical components can be fabricated; as for lithographically-defined structures, each component can be tailored to optimize properties such as quantum coherence. Here, we demonstrate what is perhaps the most potent advantage of molecular spin qubits, the scalability of quantum information processing structures using bottom-up chemical self-assembly. Using $Cr_7Ni$ spin qubit building blocks, we have constructed several families of two-qubit molecular structures with a range of linking strategies. For each family, long coherence times are preserved, and we demonstrate control over the inter-qubit quantum interactions that can be used to mediate two-qubit quantum gates.**

*Introduction*

An information processing device whose elements are capable of storing and processing quantum superposition states (a quantum computer) would support algorithms for useful tasks such as searching[1] and factoring[2] that are much more efficient than the corresponding classical algorithms[3], and would allow efficient simulation of other quantum systems[4]. One of the key challenges in realizing a quantum computer lies in identifying a physical system that hosts quantum states sufficiently coherently, and provides appropriate interactions for implementing logic operations[5]. Among the molecular spin systems that have been proposed as qubit candidates are N@$C_{60}$[6,7,8,9], organic radicals[10,11], and molecular magnets[12,13,14]. We proposed exploiting molecular magnets based on heterometallic antiferromagnetic rings[15,16]. These systems exhibit a number of favourable features supporting their application as components of a quantum computer: flexibility in their chemical composition allows control over both the total ground state spin (by modifying the hetero-atom) and the carboxylate ligands[17]; their well-defined internal magnetic excitations may offer mechanisms for efficient single-qubit manipulations[15]; and the ground state spin is highly coherent[18], particularly when the chemical structure is optimised[19].

In the context of molecular spin qubits, the simplest conceivable multi-qubit structure is a molecular dimer. This observation has motivated various efforts to synthesise dimers including, for example, N@$C_{60}$ -- N@$C_{60}$[20,21], radical -- radical[10,11], N@$C_{60}$ -- molecular magnet[22], and molecular magnet -- molecular magnet[23,24,25,26]. However, the design of a dimer specifically to host two-qubit experiments should take account of the importance of three key time scales relative to one another: $T_2$, the individual qubit phase relaxation time, must be longest; $h/J$, which is characteristic of the duration of two-qubit gates (where $J$ is the inter-qubit interaction energy and $h$ is Planck's constant) should be intermediate; and the single-qubit manipulation time should be the shortest. In practice, phase relaxation times in heterometallic antiferromagnetic rings are in the 1 to 10 $\mu$s range at low temperatures[18,19] and they can be manipulated in a typical pulsed electron spin resonance (ESR) apparatus on the 10 ns timescale; thus an interaction offering $h/J$ in the 100 ns range could be exploited, for example, in a multi-qubit experiment to generate controlled entanglement. In this Article we report the synthesis of two families of dimers of the antiferromagnetic ring $Cr_7Ni$ and pulsed electron spin resonance (ESR) experiments probing the spin coherence times and the intra-dimer magnetic interactions. We demonstrate that the modular nature of our synthetic approach provides independent control of the molecular nanomagnet components and of the chemical coupling between them. This allows us the flexibility to optimise the physical properties of the dimers with respect to the three key time scales.

*Synthesis of multi-qubit molecular structures*

The molecular structures of the dimers that we studied are shown in Figure 1. The monomer components are based on $Cr_7Ni$ rings in which intra-ring nearest-neighbour antiferromagnetic coupling gives rise to a well-defined $S$=1/2 ground state. Each dimer family employs a different variety of $Cr_7Ni$ ring, allowing very different approaches to chemical dimerisation.

The first family (**1A**, **1C** and **1D** in Figure 1) is a collection of hybrid [3]rotaxanes[27]. In each of these compounds two $Cr_7Ni$ rings are threaded by a rigid organic molecule. There is no covalent bonding between the two rings and the magnetic interaction is expected to be purely dipolar (through space), and modulated by both the length of the threading molecule and the orientation of the dimer with respect to an external magnetic field.

The [3]rotaxane structures, are synthesised by growing the heterometallic rings around rigid thread molecules that contain amine-binding sites. Initially we studied two closely related threads: [R-$CH_2CH_2$-NH-$CH_2$-$(C_6H_4)_3$-$CH_2$-NH-$CH_2CH_2$-R] (R = Ph, **Th1**; R = $CH_2CHMe_2$, **Th2**). For both threads there are three aromatic rings between the two amines, and they differ only in the stopper on the thread. The reaction to make the [3]rotaxanes involves adding the organic thread to a solution of hydrated chromium trifluoride dissolved in pivalic acid ($HO_2C^tBu$); the heterometallic rings then grow around each ammonium cation and the ring is completed by addition of a nickel precursor. We made four related [3]rotaxanes (see Figure 1 and Supplementary Information): {[$H_2$**Th1**][$Cr_7NiF_8(O_2C^tBu)_{16}$]$_2$} **1A**; {[$H_2$**Th2**][$Cr_7NiF_8(O_2C^tBu)_{16}$]$_2$} **1B**; {[$H_2$**Th2**][$Cr_7NiF_8(O_2C$*d*-$^tBu)_{16}$]$_2$} **1Bd** (with the pivalic acid perdeuterated); {[$H_2$**Th1**][$Cr_7NiF_8(O_2CAd)_{16}$]$_2$} **1C** ($HO_2Cad$ = 1-adamantanecarboxylic acid) in yields between 17 and 35%.

In the [3]rotaxanes the two $Cr_7Ni$ rings each contain an octagonal array of seven chromium(III) sites and one nickel(II) site, with each edge of the octagon bridged by a single fluoride and two carboxylates; these molecules are green in colour. The dimer structures can essentially be regarded as an octagonal prism of metal sites. The fluorides lie inside the octagon and form H-bonds to the ammonium of the thread. The distance between the $Cr_7Ni$ ring centroids is 1.64 nm, while the longest metal--metal distance between rings is 1.94 nm. The ring cavity is sufficiently small that a phenyl- or *iso*-propyl group can act as a stopper for the rotaxane. In the crystal structure the nickel site is disordered about the eight metal sites of the octagon. We made compound **1Bd** because we have found that deuterating ligands in $Cr_7Ni$ monomers can significantly prolong the spin coherence times[18]. Similarly spin coherence times in isolated rings can be improved by reducing the mobility of nuclear spins in the ligands[19], leading us to prepare **1C**, which contains carboxylate ligands with rigid adamantyl groups.

Attempts to increase the inter-ring separation by adding aromatic rings to the linker were limited by the low solubility of longer organic chains containing rigid aromatic groups. Instead, we made a new [2]rotaxane by growing the ring around Ph-$CH_2NH_2$-$CH_2$-$C_6H_4$-Py **Th3** (Py = pyridyl), giving {[H**Th3**][$Cr_7NiF_8(O_2C^tBu)_{16}$]} (Figure S2); this [2]rotaxane contains a pyridyl group at the end of the thread and two of these are then coordinated to the axial position of a dimetal tetracarboxylate compound, in this case [$Rh_2(O_2CMe)_4$]. This gives a new route to [3]rotaxanes, and results in {[H**Th3**][$Cr_7NiF_8(O_2C^tBu)_{16}$]}$_2$[$Rh_2(O_2CMe)_4$] **1D** (Figure 1c), where there is a distance of 2.5 nm between the $Cr_7Ni$ ring centroids of the [3]rotaxane.

The second family of dimers (**2A**, **2B** and **2C** in Figure 1) involves a chiral $Cr_7Ni$ ring, [$Cr_7NiF_3(Etglu)(O_2C^tBu)_{15}(H_2O)$] (where $H_5$Etglu = N-ethyl-D-glucamine)[28]. This structure again contains an octagon of metal sites, but at the centre is a penta-deprotonated N-ethyl-D-glucamine, which bridges between Cr sites using the deprotonated alkoxide side-arms. Fluorides bridge three edges of the octagon, and there are again pivalate groups outside the ring. These molecules are purple in colour. The key difference from the rings in the first family is that there is a substitutionally labile water molecule on the nickel(II) site, which is unique and defined crystallographically. Reacting two such rings with, for example, a linear di-imine gives dimers linked via the Ni...Ni axis.

Previously we have used simple di-imines[24], producing dimers in which we can vary the inter-ring interaction through distance and through the torsion angle within the bridging group. The short di-imine linker leads to inter-ring interaction energies of the order of $h\times0.3$ GHz to $h\times10$ GHz, giving rise to large splittings in the continuous wave (CW) ESR spectrum compared to the monomer[24]. Interactions on this scale would give rise to inter-qubit gate times that are shorter than the duration of single-qubit gates, precluding the possibility of multi-qubit pulsed ESR experiments.

In order to extend the separation of the covalently-bound ring dimers, we have employed boronic acid-capped clathrochelate complexes[29], which have previously been used as scaffolds to create bipyridyl linkers varying in length from 1.5 to 5.4 nm. The three dimers in this family all contain Fe(II) (low-spin and diamagnetic in this trigonal prismatic coordination environment) as the divalent metal within a capped tris(dioxime) macrobicycle. We synthesised the clathrochelates as described elsewhere[29] and mixed them with $[Cr_7NiF_3(Etglu)(O_2C^tBu)_{15}(H_2O)]$, leading to compounds $\{[Cr_7NiF_3(Etglu)(O_2C^tBu)_{15}]_2L1\}$ **2A**, $\{[Cr_7NiF_3(Etglu)(O_2C^tBu)_{15}]_2L2\}$ **2B** and $\{[Cr_7NiF_3(Etglu)(O_2C^tBu)_{15}]_2L3\}$ **2C**, shown in Figure 1. Compounds **2A** and **2C** crystallised well, allowing a full structure determination. Crystals of **2B** were weakly diffracting and only the metal sites could be determined, nevertheless allowing us to estimate the distance between the Ni sites on the heterometallic rings. The shortest metal...metal contacts in this family are the Ni...Ni distances at 1.89 nm, 2.71 nm, and 3.07 nm for compounds **2A**, **2B**, and **2C** respectively. The longest Cr...Cr distances between rings are far greater at 2.24, 3.24 and 3.55 nm respectively.

*Electron spin resonance experiments*

Figure 2 shows X-band (approximately 9.5 GHz) CW ESR spectra of compounds **1A** and **2A** and EasySpin[30] simulations, from which the spin Hamiltonian parameters may be extracted (see Figure 2 caption). Within each dimer family, the spectra are very similar; spectra for **1B**, **1C**, **1D**, **2B**, and **2C** are shown in Supplementary Information (Figure S1). For both **1A** and **2A**, the spectra are characteristic of a spin-1/2 exhibiting an approximately axially anisotropic *g*-factor. The X-band spectrum of **2A** shows more structure on the low-field edge; we believe that this due to partial resolution of hyperfine structure. In both, there is a broadening arising from unresolved hyperfine interactions.

In both compounds the CW spectra are similar to the spectra of the corresponding monomers[18,31], indicating that in these compounds the inter-ring magnetic interaction is small compared to the ESR linewidth in energy. This is consistent with the requirement that two-qubit gates should take longer than single-qubit manipulations, a first indication that these dimers are more suitable for quantum information processing experiments than earlier heterometallic ring dimers[24]. X-band pulsed ESR experiments yield spin coherence times, $T_m$ (given in Table 1) that are broadly consistent with the times measured for monomers[19], offering reassurance that the formation of complexes and dimers is not intrinsically detrimental to the quantum phase coherence.

The form of the CW spectra imposes an upper bound on the intra-dimer magnetic interaction, but to measure this interaction precisely in this regime requires coherent methods; double electron -- electron resonance (DEER, also known as pulsed electron double resonance, PELDOR)[32,33,34] is the established method for so doing. Although originally applied to inorganic model systems[32], in recent years, DEER has been applied extensively to biological systems[35]. By modifying pairs of sites in proteins with spin labels such as nitronyl nitroxides, DEER can be used to measure the dipolar interaction, and therefore the distance, between the labels[36].

DEER techniques exploit the fact that, if there is a magnetic interaction between two spin centres, the coherent precession rate of one of the centres is modified by an inversion of the spin state of the other centre. In practice this is achieved through pulsed ESR sequences using two frequencies, such as the standard four-pulse DEER sequence shown in Figure 3(a)[33]. The first two pulses, $\pi/2$ -- $\tau_1$ -- $\pi$ at frequency $\nu_1$, set up a Hahn echo[34] at a time $2\tau_1$ on one of the spins. As this spin continues to precess, a "pump" $\pi$ pulse at a second frequency $\nu_2$ at time $T$ (relative to the $\nu_1$ echo) inverts the second spin, thus modifying the effective magnetic field (and therefore the precession rate) experienced by the first spin. A subsequent refocusing $\pi$ pulse at frequency $\nu_1$ a time $\tau_2$ after the first echo generates a second echo at a time $2\tau_1+2\tau_2$, with an amplitude that depends on the time at which the $\nu_2$ inversion pulse was applied and on the strength of the interaction between the two spins. Observing this echo as a function of the time, $T$, of the pump pulse reveals oscillations at the frequency corresponding to the magnetic interaction energy between the two centres. A three-pulse variant, shown in Figure 3(b)[32], has the advantage of being shorter and having fewer pulses, thus enhancing the amplitude of the measured echo, particularly for systems in which the spin coherence times are short. The disadvantage is that for short times $T$, the $\nu_2$ pump pulse overlaps the coherence-generating $\nu_1$ $\pi/2$-pulse, distorting the spectra around the time $T = 0$ [33].

This description of DEER depends on some heterogeneity between the two spin centres in the molecule under study, so that the frequencies $\nu_1$ and $\nu_2$ are resonant with different centres. We performed DEER experiments at low temperatures (2.5 K, to maximise spin coherence times) on dilute frozen solutions of dimers (0.1 -- 0.2 mM, to minimise inter-dimer dipolar interactions). In this case differences between the two heterometallic rings may arise from conformational flexibility. In the first family, the [3]rotaxane dimers, there may be variation in the position of the Ni centre between the two rings, or some slight bending of the stiff threading molecule. In the second family, the covalently bound dimers, the rings could rotate about the axis defined by the linker. In either case, the distortions of individual rings (giving rise to *g*-strains) or variations in configurations of hyperfine-coupled nuclei (resulting in variations of the effective local magnetic field) offer asymmetry between the two rings within a dimer.

The CW spectra in Figure 1 extend over a magnetic field range of about 30 mT at X-band (about 9.5 GHz), corresponding to an absorption spectrum at a fixed magnetic field covering about 750 MHz, as plotted in Figure 3(c). Intervals within this energy range can be identified with orientationally selected sub-populations of rings. Thus, microwave pulses at different frequencies within the spectrum excite orientations selectively, shown as coloured regions on spheres in Figure 3(d). We anticipate that, notwithstanding a degree of flexibility, the orientations of two rings within a single dimer should be reasonably well correlated. With this in mind, we choose to separate $\nu_1$ and $\nu_2$ by a frequency that is small compared to the total width of the spectrum, so that they excite neighbouring orientational sub-populations of rings.

In practice, the response of the microwave resonator used in the experiment is not uniform over the frequency range of the absorption line. Instead we used fixed frequencies for $\nu_1$ and $\nu_2$, and we achieved orientation selection by adjusting the external magnetic field to bring the appropriate part of the absorption spectrum into resonance with the applied pulses. We used detection pulses (at frequency $\nu_1$) of 40 ns and pump pulses ($\nu_2$) of 24 ns, separated in frequency by 80 MHz.

*Results*

Figure 4 shows typical DEER spectra taken on compound **1A**. The left panel shows the amplitude of the final echo as a function of the time, *T*, of the pump pulse for three different orientation selections. Open circles are raw data; the solid curves are filtered using standard tools (the Matlab package DeerAnalysis[37]) to smooth the data and to remove oscillations (electron spin echo envelope modulation, or ESEEM[18,34]) associated with couplings to proton nuclear moments. To the right of each trace, a spherical intensity map indicates the range of orientations probed at the given applied magnetic field and experimental detection frequency. The right panel shows the Fourier transforms of the time-domain data from the left panels (both raw data, open circles, and filtered data, solid lines).

The dependence of the DEER signal on magnetic field (or, equivalently, orientational sub-population) is strong, owing to several contributory factors. First, the inter-ring dipolar interaction depends on the orientation, $\theta$, of the dimer inter-ring axis with respect to the magnetic field as ($3\cos^2\theta - 1$). It therefore varies significantly as a function of the dimer orientation and passes through zero at the magic angle, $\theta \approx 54°$. Second, the magnitude by which the detected echo is modulated in the DEER experiment depends on the proportion of dimers excited initially by $\nu_1$ that is also excited, at the other end, by $\nu_2$.

In a magnetic field of 377.84 mT (lowest trace, blue), dimers whose inter-spin axes are perpendicular to the field are predominantly excited. Within this sub-population the range of inter-ring magnetic dipolar interactions is strongly peaked around a single value, giving rise to clearly-defined oscillations with a period of about 155 ns. The Fourier transform, which represents the distribution of coupling strengths within the selected orientational sub-population, is correspondingly sharply peaked at about 6.5 MHz. The first minimum in the time-domain data, occurring at about 77 ns, is the duration for inversion of the state of one of the qubits under the influence of the second. This evolution of the state of one qubit conditional on the state of the other is exactly the kind of physical interaction that allows for implementation of multi-qubit quantum logic; in our case, the time to the first minimum corresponds to the duration of a two-qubit conditional phase gate. We note that this time satisfies the criteria that we identified above for two-qubit devices: the two-qubit gate time (of order 100 ns) lies between the single-qubit gate time (of order 10 ns) and the phase relaxation time (in the microsecond range).

The sub-population excited at a magnetic field of 385.84 mT (middle trace, green) is comprised of dimers whose inter-ring axes lie at orientations peaked in between parallel and perpendicular to the external magnetic field. The distribution of orientations includes a significant proportion close to the magic angle at which the dipolar interaction strength goes through zero, resulting in a broad distribution of interaction strengths and an almost monotonic dependence of the time domain DEER signal on *T*.

At 390.84 mT, $\nu_1$ excites principally rings whose planes are perpendicular to the magnetic field (i.e. those whose effective *g*-factor is close to $g_{//}$). Deviations from this orientation lead to a greater contribution to the effective *g*-factor from the larger $g_{\perp}$, and therefore to a higher resonant frequency. Thus, in dimers that are orientationally selected by the excitation of one ring at frequency $\nu_1$, $\nu_2$ (which is lower in frequency than $\nu_1$) is comparatively unlikely to excite the other ring. This gives rise directly to the comparatively weak modulation in the time-domain trace at 390.84 mT. The inter-spin axes of this sub-population of dimers are aligned along the magnetic field direction, maximising the orientation-dependence of the dipolar interaction (i.e., $\theta \approx 0°$). Correspondingly, the time-domain signal evolves comparatively rapidly at short times and the Fourier transform contains components at high frequencies.

Figure 5 shows the equivalent DEER data for a sample from the second family, **2A**. As for sample **1A**, the DEER signal reveals well-defined oscillations and a strongly peaked distribution of inter-ring couplings for populations of dimers whose axes are close to perpendicular to the magnetic field, and broader coupling distributions and weaker modulations for populations at other orientations. However, compared with **1A**, **2A** exhibits coupling strengths that are smaller by a factor of about two (the first minimum in the time-domain data at 380.84 mT occurs at about 157 ns; the peak in the Fourier transform is at 3.3 MHz). This is broadly accounted for by the larger separation of the rings in **2A** (which is somewhat extended over the Ni...Ni distance of 1.89 nm by the rotational degree of freedom of the rings about the linker axis) than in **1A** (the centroids of whose rings are separated by 1.6 nm), and the fact that the dipolar interaction varies with separation, *r*, as $1/r^3$.

These trends are reflected in the DEER data for all of the compounds shown in Figure 1; the time-domain data and Fourier transforms for compounds other than **1A** and **2A** are presented in the Supplementary Information Figures S2 to S6, and the two-qubit gate times are listed in Table 1. There is a direct correlation between the interaction strength and the inter-ring separation, with larger separations giving rise to weaker interactions. The two-qubit gate times reach about 550 ns for the longest dimers. The interactions are weaker in the the second family than in the first, owing to the extra degree of freedom in the structure: rings that are attached to the linker through a single point on the rim of the ring are free to rotate about the linker axis, allowing a greater inter-ring separation.

We note also that our strategy allows for enhancement of the spin coherence times independently of the inter-ring interaction strength. In earlier studies of monomers[19], we found that deuteration and immobilization of nuclear spins in the structure significantly improve the coherence. Thus, in compounds **1A**, **1Bd** and **1C**, while the similar linkers ensure that $h/J$ is similar for each, deuteration (**1Bd**) and confinement of hydrogen nuclei to rigid adamantyl groups (**1C**) result in extensions of the coherence times by factors of about four and three respectively.

*Discussion*

From the data presented here, we may draw several conclusions supporting the assertion that molecular spin systems may form useful components in future quantum technologies. Molecular magnets offer building blocks which, through supramolecular chemistry, can be assembled into multi-qubit structures. Identifying strict criteria for the relative timescales for single-qubit manipulations, two-qubit interactions and phase memory times, we have designed several families of structures to optimise the prospects of hosting two-qubit devices. For example, in compound **1A**, the phase coherence time of about 800 ns significantly exceeds the two-qubit gate time of 77 ns, and both are much longer than the typical single-qubit manipulation time of about 10 ns. Thus we build upon the wide range of proposals to exploit molecular nanomagnets in quantum information by demonstrating for the first time coherent manipulation of dimers of such species, including multi-frequency experiments probing the inter-qubit interaction, and by showing that we can chemically tune the interaction strength to meet architectural requirements. This prepares the path for experiments demonstrating controlled entanglement operations (as opposed to the observation of entanglement as an inherent property of a singlet ground state of two coupled spins[38]), and other two-qubit manipulations.

*References*

1. Grover, L.K. Quantum mechanics helps in searching for a needle in a haystack. *Phys. Rev. Lett.* **79**, 325–328 (1997).

2. Shor, P.W. Polynomial-time algorithms for prime factorization and discrete logarithms on a quantum computer. *SIAM J. Comput.*, **26(5)**, 1484–1509 (1997).
3. Nielsen M.A. & Chuang I.L. *Quantum computation and quantum information* (Cambridge University Press, Cambridge 2000).
4. Feynman, R.P. Simulating physics with computers. *International Journal of Theoretical Physics* **21**, 467–488 (1982).
5. DiVincenzo, D.P. The physical implementation of quantum computation. *Fortschritte der Physik* **48**, 771–783 (2000).
6. Almeida Murphy, T., Pawlik, Th., Weidinger, A., Höhne, M., Alcala, R. & Paeth, J.-M. Observation of atomlike nitrogen in nitrogen-implanted $C_{60}$. *Phys. Rev. Lett.* **77**, 1075–1078 (1996)
7. Harneit, W. Fullerene-based electron spin quantum computer. *Phys. Rev. A* **65**, 032322 (2002).
8. Benjamin, S.C., Ardavan, A. Briggs, G.A.D., Britz, D.A., Gunlyke, D., Jefferson, J. *et al.* Towards a fullerene-based quantum computer. *J. Phys.: Condens. Matter* **18**, S867 (2006).
9. Morton J.J.L., Tyryshkin, A.M., Ardavan, A., Benjamin, S.C., Porfyrakis, K. *et al.* Bang-bang control of fullerene qubits using ultrafast phase gates. *Nature Physics* **2**, 40–43 (2006)
10. Sato, K., Nakazawa, S., Rahimi, R., Ise, T., Nishida, S., Yoshino, T. *et al.* Molecular electron-spin quantum computers and quantum information processing: pulse-based electron magnetic resonance spin technology applied to matter spin-qubits. *J. Mater. Chem.* **19**, 3739–3754 (2009).
11. Ayabe, K., Sato, K., Nishida, S., Ise, T., Nakazawa, S., Sugisaki, K. *et al*. Pulsed electron spin nutation spectroscopy of weakly exchange-coupled biradicals: a general theoretical approach and determination of the spin dipolar interaction. *Phys. Chem. Chem. Phys.* **14**, 9137–9148 (2012).
12. Leuenberger M.N. & Loss, D. Quantum computing in molecular magnets. *Nature* **410**, 789–793 (2001).
13. Tejada, J., Chudnovsky, E.M., del Barco, E., Hernandez, J.M. & Spiller T.P. Magnetic qubits as hardware for quantum computers. *Nanotechnology* **12**, 181 (2001).
14. Meier, F., Levy, J., Loss, D. Quantum computing with spin cluster qubits. *Phys. Rev. Lett.* **90**, 047901 (2003)

15. Troiani, F., Ghirri, A., Affronte, M., Carretta, S., Santini, P., Amoretti, G. *et al.* Molecular engineering of antiferromagnetic rings for quantum computation. *Phys. Rev. Lett.* **94**, 207208 (2005)
16. Ardavan, A. & Blundell S.J. Storing quantum information in chemically engineered nanoscale magnets. *J. Mater. Chem.*, **19**, 1754–1760 (2009).
17. Larsen, F.K., McInnes, E.J.L., El Mkami, H., Overgaard, J., Piligos, S., Rajaraman, G. *et al.* Synthesis and characterization of heterometallic {$Cr_7$M} wheels. *Angew. Chem. Int. Ed.* **42**, 101–105 (2003).
18. Ardavan, A., Rival, O., Morton, J.J.L., Blundell, S.J., Tyryshkin, A.M., Timco G.A. *et al.* Will spin relaxation times in molecular magnets permit quantum information processing? *Phys. Rev. Lett.* **98**, 057201 (2007).
19. Wedge, C.J., Timco, G.A., Spielberg, E.T., George R.E., Tuna, F., Rigby, S. *et al.* Chemical engineering of molecular qubits. *Phys. Rev. Lett.* **108**, 107204 (2012).
20. Farrington, B.J., Jevric, M., Rance, G.A., Ardavan, A., Khlobystov, A.N., Briggs, G.A.D. *et al.* Chemistry at the nanoscale: Synthesis of an N@$C_{60}$–N@$C_{60}$ endohedral fullerene dimer. *Angew. Chem. Int. Ed.* **51**, 3587–3590 (2012).
21. Plant, S.R., Jevric, M., Morton, J.J.L., Ardavan, A., Khlobystov, A.N., Briggs, G.A.D. *et al*. A two-step approach to the synthesis of N@$C_{60}$ fullerene dimers for molecular qubits. *Chem. Sci.* **4**, 2971–2975 (2013).
22. Kaminski, D. *et al.*, in preparation.
23. Hill, S., Edwards, R.S., Aliaga-Alcalde, N. & Christou, G., Quantum coherence in an exchange-coupled dimer of single-molecule magnets. *Science* **302**, 1015–1018.
24. Faust, T.B., Bellini, V., Candini, A., Carretta, S., Lorusso, G., Allan, D.R. *et al*. Chemical control of spin propagation between heterometallic rings. *Chem. Eur. J.* **17**, 14020–14030 (2011).
25. Luis, F., Repollés, A., Martínez-Pérez, M.J., Aguilà, D., Roubeau, O., Zueco, D. *et al.* Molecular prototypes for spin-based CNOT and SWAP quantum gates. *Phys. Rev. Lett.* **107**, 117203 (2011).
26. Aguila, D., Barrios, L.A., Velasco, V., Roubeau, O., Repollés, A., Alonso, P.J. *et al.* Heterodimetallic [LnLn′] lanthanide complexes: toward a chemical design of two-qubit molecular spin quantum gates. *J. Am. Chem. Soc.* **136**, 14215–14222 (2014).
27. Lee, C.-F. Leigh, D.A., Pritchard, R.G., Schultz, D., Teat, S.J., Timco, G.A. *et al.* Hybrid organic–inorganic rotaxanes and molecular shuttles. *Nature* **458**, 314–318 (2009).

28. Garlatti, E., Albring, M.A., Baker, M.L., Docherty, R.J., Mutka, H., Guidi, T. *et al.* A detailed study of the magnetism of chiral {$Cr_7M$} rings: an investigation into parameterization and transferability of parameters. *J. Amer. Chem. Soc.* **136**, 9763–9772 (2014).
29. Wise, M.D., Ruggi, A., Pascu, M., Scopelliti, R. & Severin, K. *Chem. Sci.* **4**, 1658–1662 (2013).
30. Stoll, S. & Schweiger, A. EasySpin, a comprehensive software package for spectral simulation and analysis in EPR. *Journal of Magnetic Resonance* **178**, 42–55 (2006).
31. Timco, G.A., McInnes, E.J.L., Pritchard, R.G., Tuna, F. & Winpenny R.E.P. *Angew. Chem. Int. Ed.* 47, 9681–9684 (2008).
32. Milov, A.D., Ponomarev, A.B. & Tsvetkov, Yu. Electron-electron double resonance in electron spin echo: model biradical systems and the sensitized photolysis of decalin. *Chem. Phys. Lett.* **110**, 67–72 (1984).
33. Pannier, M., Veit, S., Godt, A., Jeschke, G. & Spiess, H.W. Dead-time free measurement of dipole–dipole interactions between electron spins. *Journal of Magnetic Resonance* **142**, 331–340 (2000).
34. Schweiger, A. & Jeschke G. *Principles of pulse electron paramagnetic resonance* (Oxford University Press, Oxford, 2001).
35. Schiemann, O., & Prisner, T.F. Long-range distance determinations in biomacromolecules by EPR spectroscopy. *Quarterly Reviews of Biophysics* **40**, 1–53 (2007).
36. Jeschke, G. & Polyhach, Y. Distance measurements on spin-labelled biomacromolecules by pulsed electron paramagnetic resonance. *Phys. Chem. Chem. Phys.* **9**, 1895–1910 (2007).
37. Jeschke, G., Chechik, V., Ionita, P., Godt, A., Zimmermann, H., Banham, J. *et al.* DeerAnalysis2006—a comprehensive software package for analyzing pulsed ELDOR data. *Appl. Magn. Reson.* **30**, 473–498 (2006).
38. Candini, A., Lorusso, G., Troiani, F., Ghirri, A., Carretta, S., Santini, P. *et al.* Entanglement in supramolecular spin systems of two weakly coupled antiferromagnetic rings (purple-$Cr_7Ni$). *Phys. Rev. Lett.* **104**, 037203 (2010).

*Acknowledgements*
This work was supported by the EPSRC through the National EPR Service and grant EP/H012613/1.

*Author contributions*

AA, EJLM, CRT and REPW formulated the project. AF and GAT prepared the compounds, using clathrochelates prepared by MDW, AR and KS. GFSW and CAM performed the X-ray structure analysis. AMB, AJF, DK, FM and FT measured ESR data. AMB and DK analysed and interpreted data presented here. AA wrote the manuscript with substantial input from EJLM and REPW.

*Competing financial interests*

The authors have no competing financial interests.

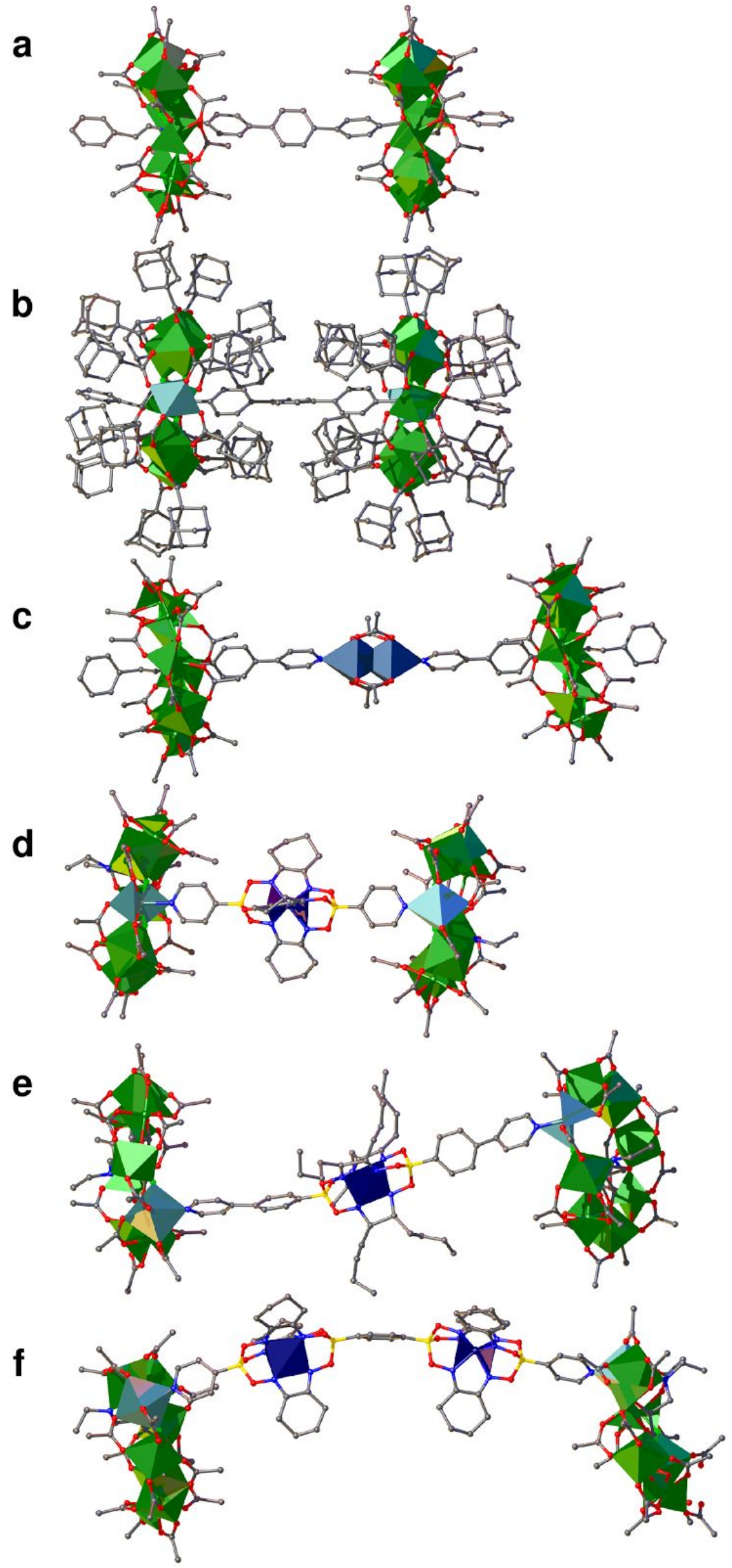

Figure 1: *The structure of the two-ring compounds in the crystal.* Cr, green octahedra; Ni, pale blue octahedra; Cu, blue octahedra; Fe, blue trigonal prisms; O, red balls; C, grey balls; N, blue balls; B, yellow balls. (a) {[$H_2$**Th1**][$Cr_7NiF_8(O_2C^tBu)_{16}]_2$} **1A** (**1B** and **1Bd** are very similar, differing only by the stopper on the thread); (b) {[$H_2$**Th1**][$Cr_7NiF_8(O_2CAd)_{16}]_2$} **1C**; (c) {[H**Th3**][$Cr_7NiF_8(O_2C^tBu)_{16}$]}$_2$[$Rh_2(O_2CMe)_4$] **1D**; (d) {[$Cr_7NiF_3(Etglu)(O_2C^tBu)_{15}]_2$L1} **2A**; (e) {[$Cr_7NiF_3(Etglu)(O_2C^tBu)_{15}]_2$L2} **2B**; (f) {[$Cr_7NiF_3(Etglu)(O_2C^tBu)_{15}]_2$L3} **2C**.

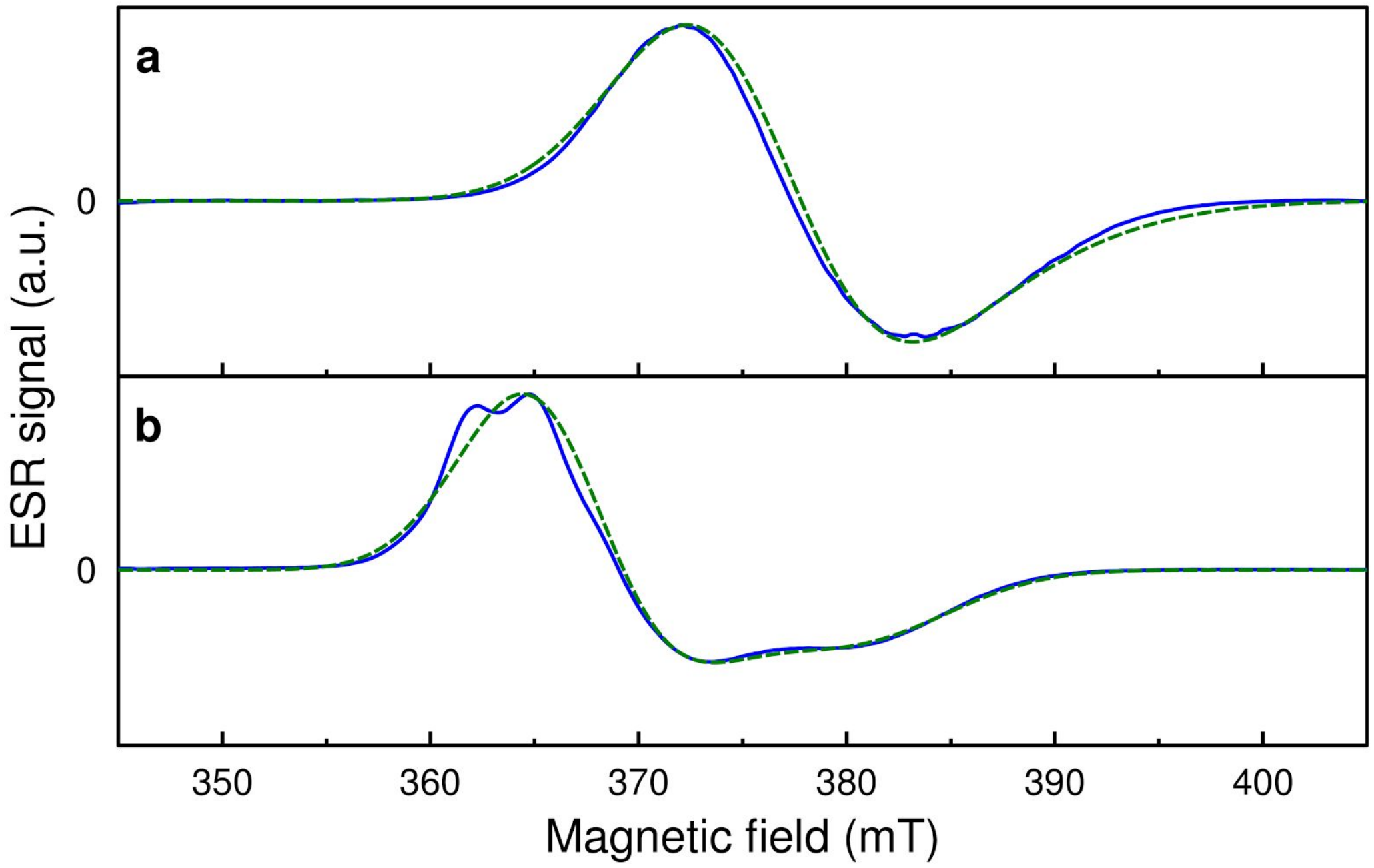

Figure 2: *CW ESR spectra of dilute frozen solutions at 2.5 K.* (a) X-band CW ESR spectrum of compound **1A** at 9.3878 GHz (blue line) and EasySpin[30] simulation (green dashed line), yielding parameters $g_{\perp}$ = 1.788 (strain, 0.02), $g_{/\!/}$ = 1.748 (strain, 0.06). A Gaussian FWHM energy broadening (H-strain) of $h$ × 250 MHz is included to account for unresolved hyperfine splittings. (b) X-band CW ESR spectrum of **2A** at 9.3892 GHz (blue line) and EasySpin simulation (green dashed line). Simulation parameters: $g_{\perp}$ = 1.83 (strain, 0.02); $g_{/\!/}$ = 1.76 (strain, 0.03); H-strain = $h$ × 200 MHz. The additional splitting, which appears in all family 2 compounds, is probably associated with partially resolved hyperfine interactions.

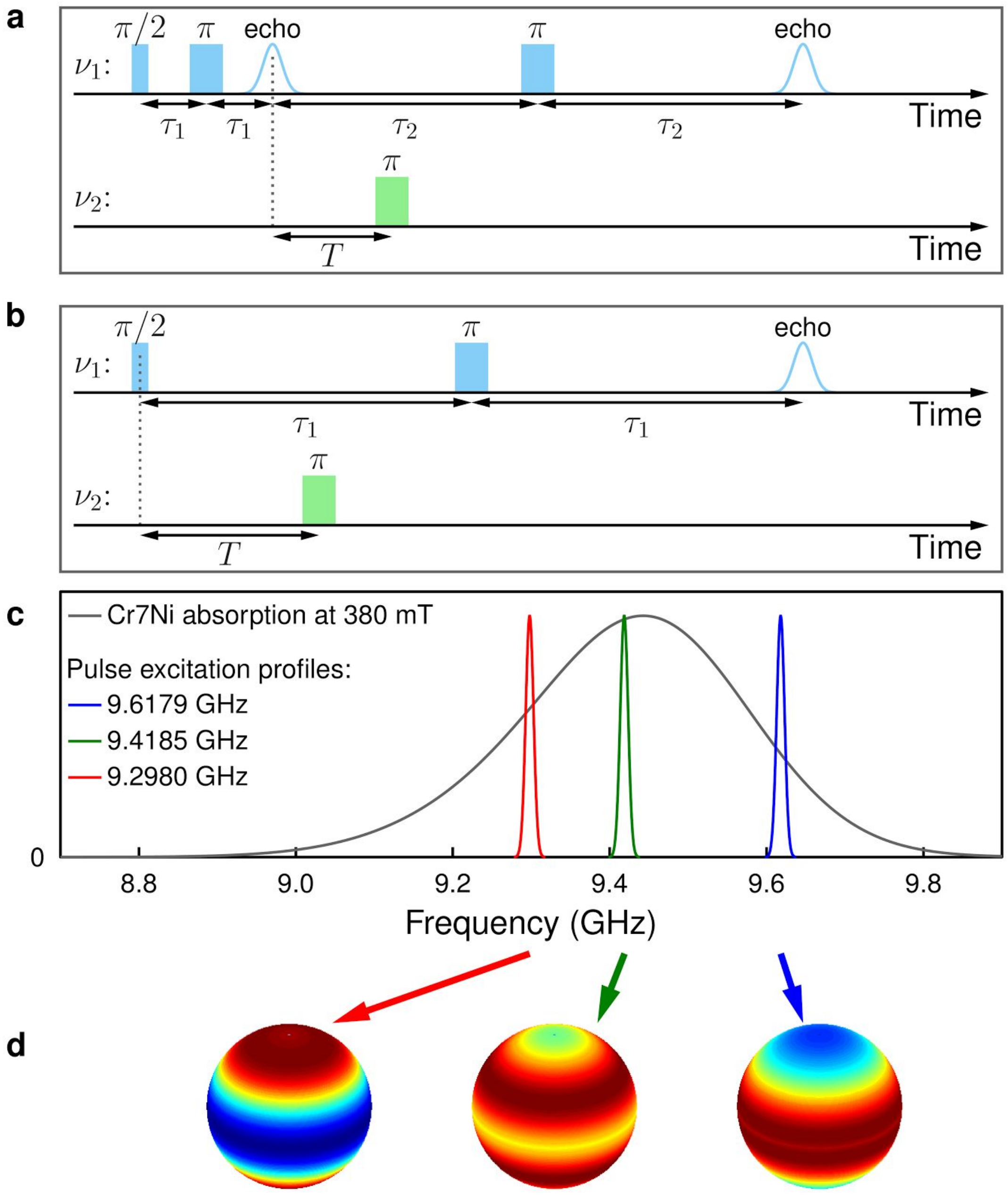

Figure 3: *Principle of the DEER experiment.* (a) 4-pulse DEER sequence. (b) 3-pulse DEER sequence. (c) ESR absorption spectrum as a function of energy (grey), and excitation profiles for pulses applied resonant with different parts of the absorption spectrum (red, green and blue). (d) Ring orientations relative to the vertical external magnetic field excited by the pulses in (c) owing to *g*-factor anisotropy and strain (red: high excitation, blue: low excitation).

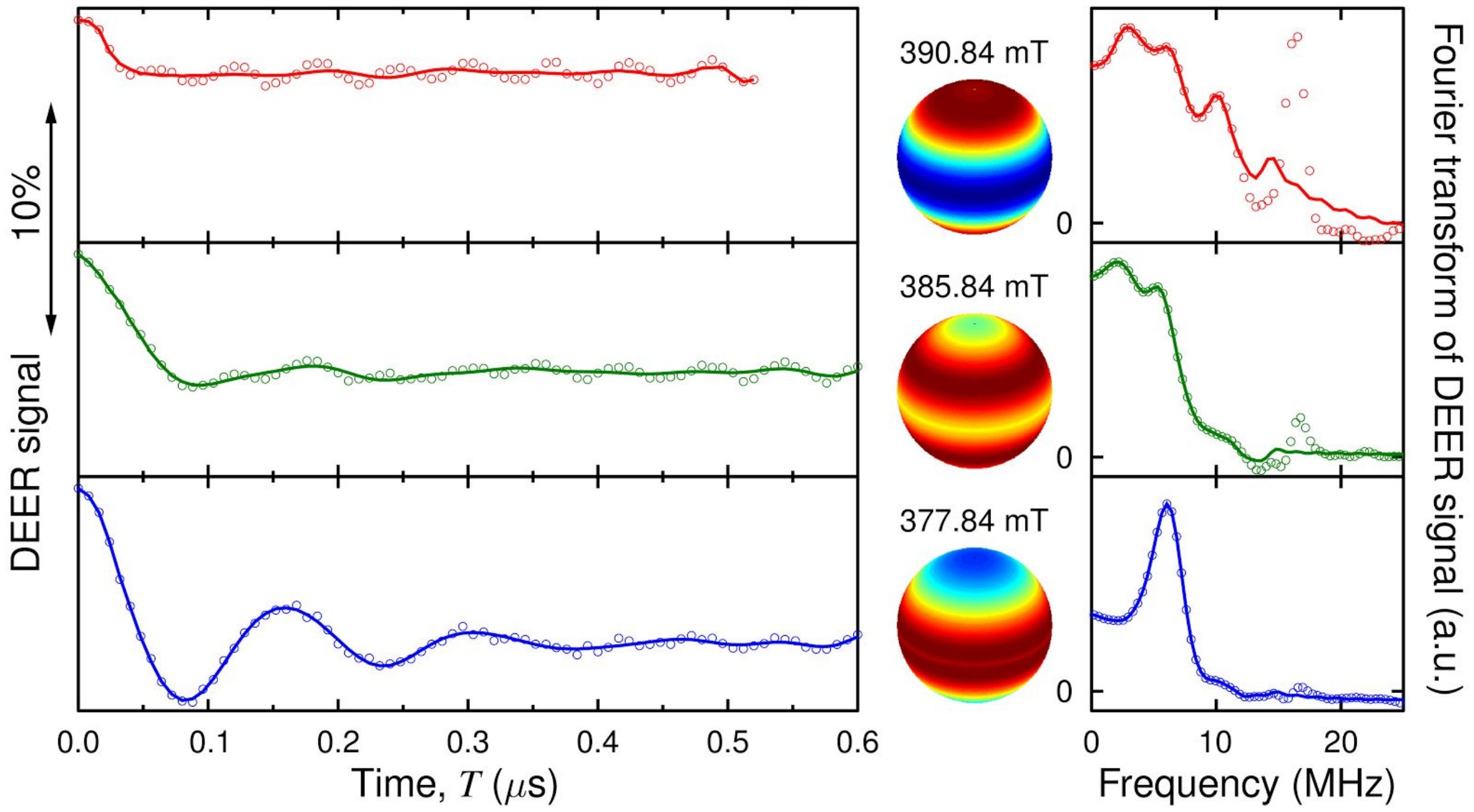

Figure 4: *DEER data for sample* ***1A***. The detection and excitation frequencies were, respectively, $\nu_1$ = 9.5632 GHz, $\nu_2 = \nu_1$ - 0.08 GHz = 9.4832 GHz. The temperature was 2.5 K. The DEER signal was recorded as a function of the time of the $\nu_1$ pump pulse (left panels) at three magnetic fields, each corresponding to an orientational sub-population indicated by the intensity maps on spheres. The Fourier transforms of the time-domain data are shown on the right. Open circles show raw data; solid lines indicate filtered data. The principal difference between the unfiltered data and the filtered data is the absence in the latter of a component at about 18 MHz arising from proton ESEEM.

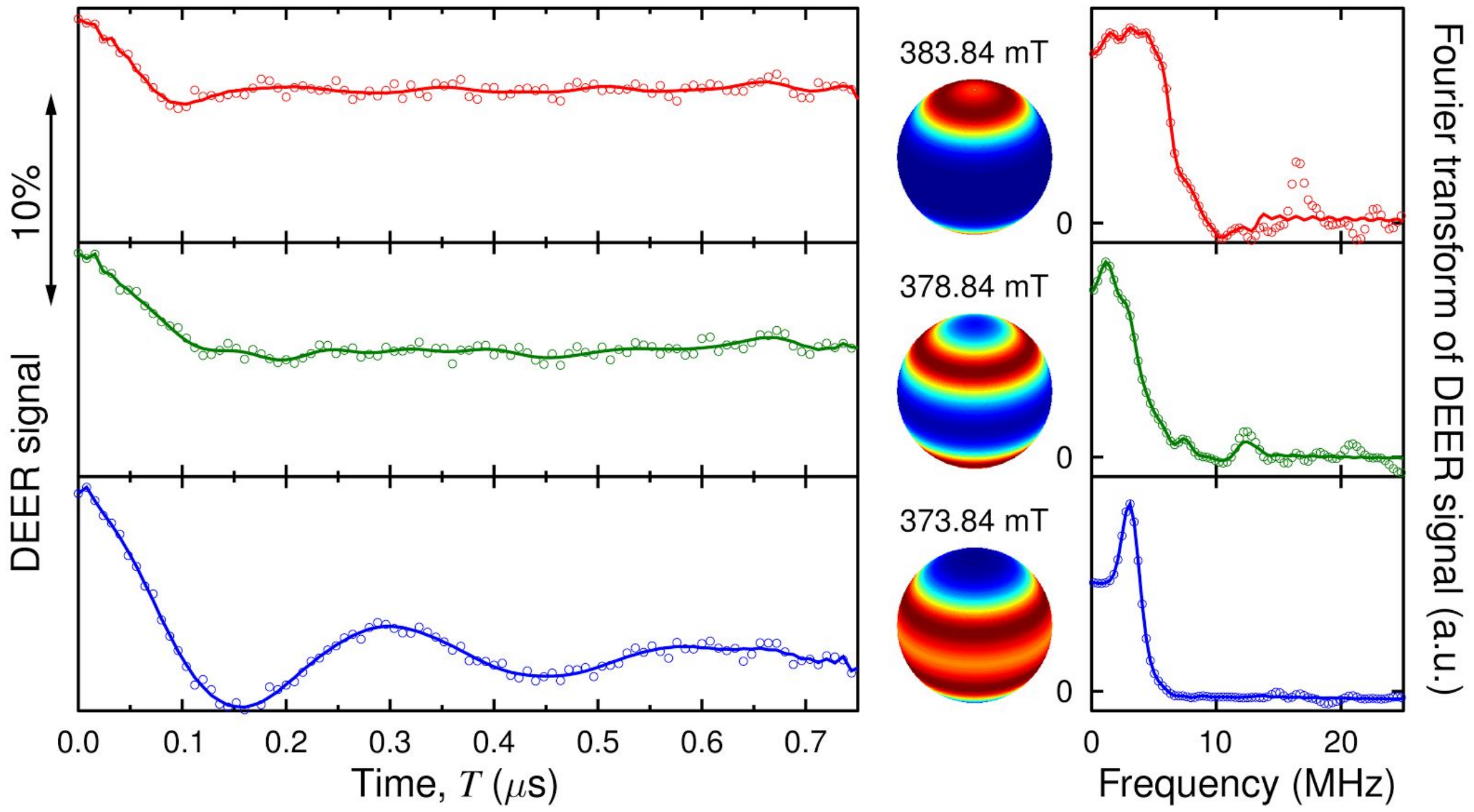

Figure 5: *DEER data for sample **2A***. The detection and excitation frequencies were, respectively, $\nu_1$ = 9.5260 GHz, $\nu_2$ = $\nu_1$ - 0.08 GHz = 9.4460 GHz.

| Compound (inter-ring distance) | $T_m$ (ns) | $x$ | Two-qubit gate time (ns) |
| --- | --- | --- | --- |
| **1A** (1.64 nm) | 801 ± 3 | 1.615 ± 0.007 | 77 |
| **1Bd** (1.64 nm) | 3239 ± 15 | 1.294 ± 0.008 | 84 |
| **1C** (1.64 nm) | 2442 ± 11 | 1.209 ± 0.007 | 82 |
| **1D** (2.50 nm) | 543 ± 8 | 1.295 ± 0.014 | 225 |
| **2A** (1.89 nm) | 608 ± 12 | 1.261 ± 0.02 | 157 |
| **2B** (2.71 nm) | 683 ± 4 | 1.446 ± 0.01 | 400 |
| **2C** (3.07 nm) | 543 ± 18 | 1.183 ± 0.03 | 550 |

Table 1: *Phase relaxation times and two-qubit gate times*. Hahn echo decays measured at 2.5 K were fitted to a phenomenological stretched exponential function of the form $A \exp[-(T_m / 2\tau)^x]$ where $2\tau$ was the total duration of the experiment, and $T_m$ and $x$ were fit parameters ($x = 1$ corresponds to a simple exponential, for which $T_m \equiv T_2$)[19]. $\pi/2$- and $\pi$-pulses were, respectively, 140 ns and 280 ns long in order to suppress proton ESEEM. Solvents used were protonated toluene (samples **1A**, **1B**, **1C**, **2B**) or deuterated toluene (**1D**, **2A**, **2C**). Concentrations were 100 μM (samples **1A**, **1B**, **1C**) or 200 μM (samples **1D**, **2A**, **2B**, **2C**). The two qubit gate time corresponds to the first minimum in the oscillatory DEER signal. (Note that for compounds **2A**, **2B**, and **2C**, the quoted inter-ring distance is a minimum; the rotational degree of freedom for rings in these structures may extend the ring separation.)

# Engineering coherent interactions in molecular nanomagnet dimers

## *Supplementary Information*

*Arzhang Ardavan,[1]* Alice M. Bowen,[1] Antonio Fernandez,[2] Alistair J. Fielding,[2] Danielle Kaminiski,[1] Fabrizio Moro,[2] Christopher A. Muryn,[2] Matthew D. Wise,[3] Albert Ruggi,[3] Eric J.L. McInnes,[2]* Kay Severin,[3] Grigore A. Timco,[2] Christiane R. Timmel,[4] Floriana Tuna,[2] George F. S. Whitehead[2] & Richard E. P. Winpenny[2]**

1. *Centre for Advanced Electron Spin Resonance, Clarendon Laboratory, Department of Physics, University of Oxford, OX1 3PU, United Kingdom*
2. *School of Chemistry & Photon Science Institute, The University of Manchester, Oxford Road, Manchester, M13 9PL, U.K.*
3. *EPFL Lausanne, Lausanne, Switzerland.*
4. *Centre for Advanced Electron Spin Resonance, Inorganic Chemistry Laboratory, South Parks Road, University of Oxford, OX1 3QR, U.K.*

## Table of Contents

# 1 Experimental and Synthetic Details

### Synthesis

### Experimental Details

Unless stated otherwise, all reagents and solvents were purchased from commercial sources and used without further purification. Compound $[Ni_2(H_2O)(O_2C^tBu)_4(HO_2C^tBu)_4]$ was prepared according the procedure reported in G. Chaboussant et al., *Dalton Trans.*, **2004**, 2758–2766**.** Terphenyl-4,4´-dicarbaldehyde was prepared according the procedure reported in M. Kozáková et al., *Synthetic Comm.* **2005**, 35, 161-167. 4-(4-Formylphenyl)pyridine was prepared according to the procedure reported in Y. You et al., *Angew.Chem. Int. Ed.* **2010**, 49, 3757 –3761**.** Compound $[Cr_7NiF_3(Etglu)(O_2C^tBu)_{15}(H_2O)]$, where $H_5Etglu$ is N-ethyl-D-glucamine ($C_8H_{14}NO_5H_5$) was prepared by literature method E. Garlatti et al., *J. Am. Chem. Soc.* **2014,** 136, 9763−9772.

The clathrochelate-based bipyridyl ligands L1, L2 and L3:

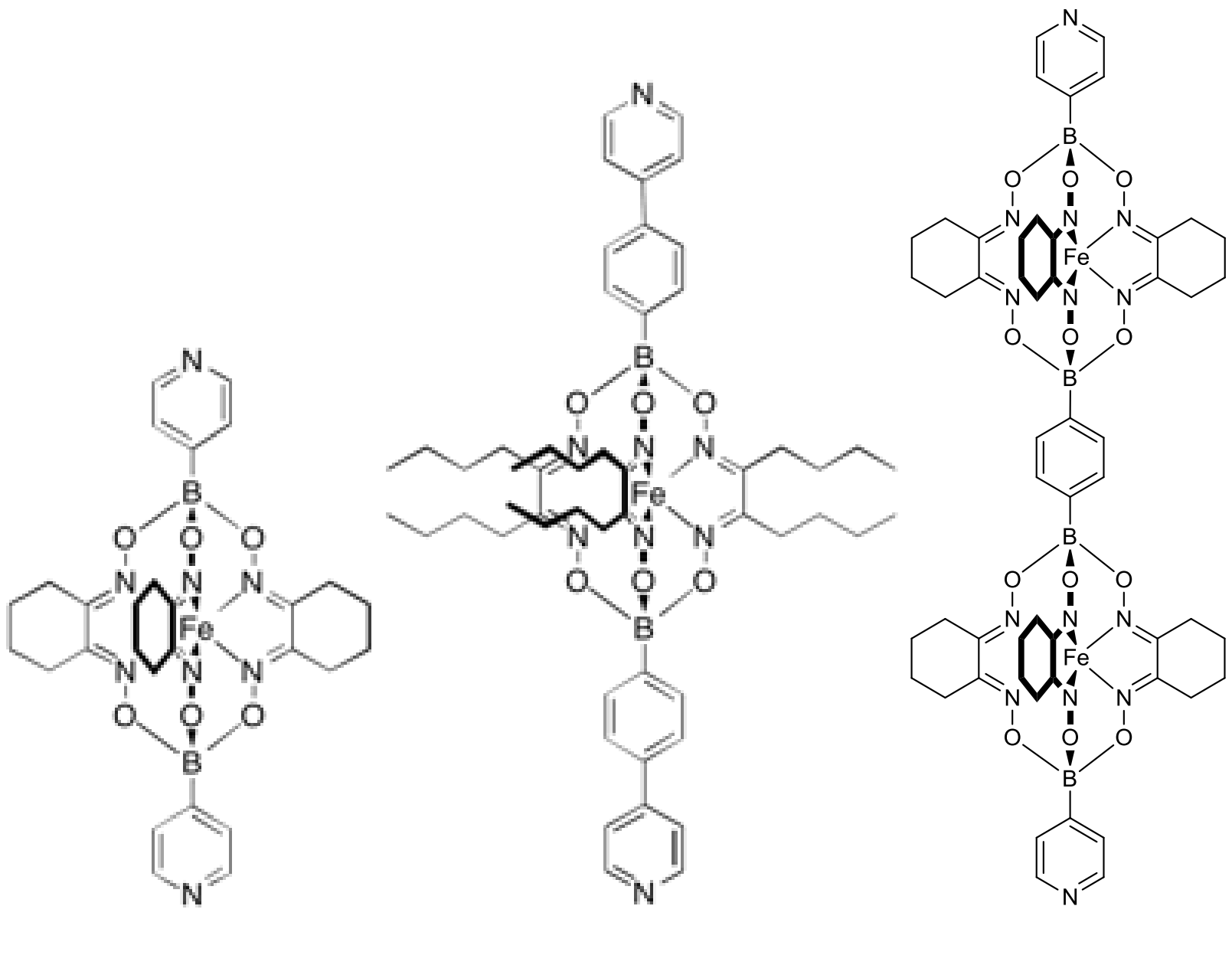

(L1 ) (L2) (L3)

were made as, or in analogy to, procedures reported in M. D. Wise, A. Ruggi, M. Pascu, R. Scopelliti. K. Severin, *Chem. Sci.*, **2013**, 4, 1658–1662.

The syntheses of the hybrid organic-inorganic rotaxanes were carried out in Erlenmeyer Teflon® FEP flasks supplied by Fisher. Column chromatography was carried out using Silica 60A (particle size 35-70 μm, Fisher, UK) as the stationary phase, and TLC was performed on precoated silica gel plates (0.25 mm thick, 60 F254, Merck, Germany) and observed under UV light. NMR spectra were recorded on Bruker AV 400, and Bruker DMX 500 instruments. Chemical shifts are reported in parts per million (ppm) from low to high frequency and referenced to the residual solvent resonance. ESI mass spectrometry and microanalysis were carried out by the services at the University of Manchester

**Synthesis of the threads:**

**Thread T1:** [Ph-$CH_2CH_2$-NH-$CH_2$-$(C_6H_4)_3$-$CH_2$-NH-$CH_2CH_2$-Ph]

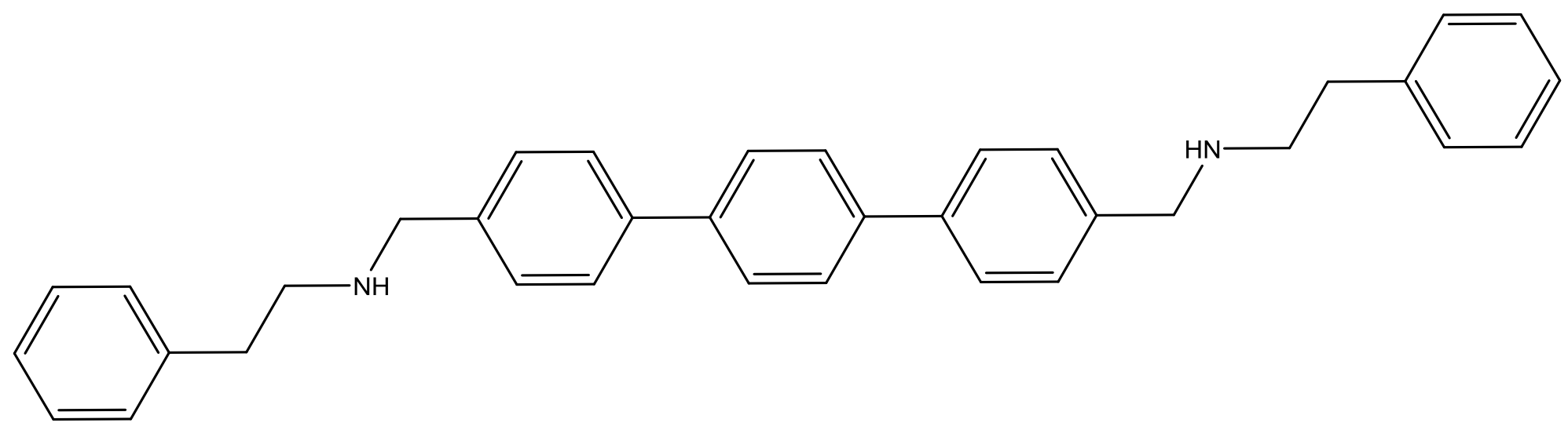

Phenethylamine (0.74 mL, 5.9 mmol) in methanol (5 mL) was added to a solution of terphenyl-4,4´-dicarbaldehyde (0.85 g, 2.9 mmol) in methanol (30 mL), and the reaction mixture was refluxed for 3 h under nitrogen, then allowed to stir at room temperature overnight. $NaBH_4$ (10 equivalents) was added and reaction mixture was stirred for 12 h under nitrogen atmosphere. The reaction was quenched with water and the solvent was evaporated. The solid was extracted with $CHCl_3$ (50 mL), washed with water and dried over anhydrous magnesium sulphate and evaporated. A light yellow liquid was obtained in 78 % yield (1.14 g). ESI-MS (sample dissolved in MeOH, run in MeOH): m/z = 497 $[M+H]^+$. $^1$H NMR (500 MHz, 293K, $CDCl_3$): δ = 1.40 (s br, 2H, NH), 2.81-2.86 (t, 4H), 2.90-2.97 (t, 4H), 3.84 (s, 4H), 7.2-7.4 (m, 10H), 7.51 (d, 4H) 7.62 (d, 4H), 7.7 (s, 4H).

**Thread T2:** [$(CH_3)_2$CH-$CH_2$-NH-$CH_2$-$(C_6H_4)_3$-$CH_2$-NH-$CH_2$-$CH(CH_3)_2$]

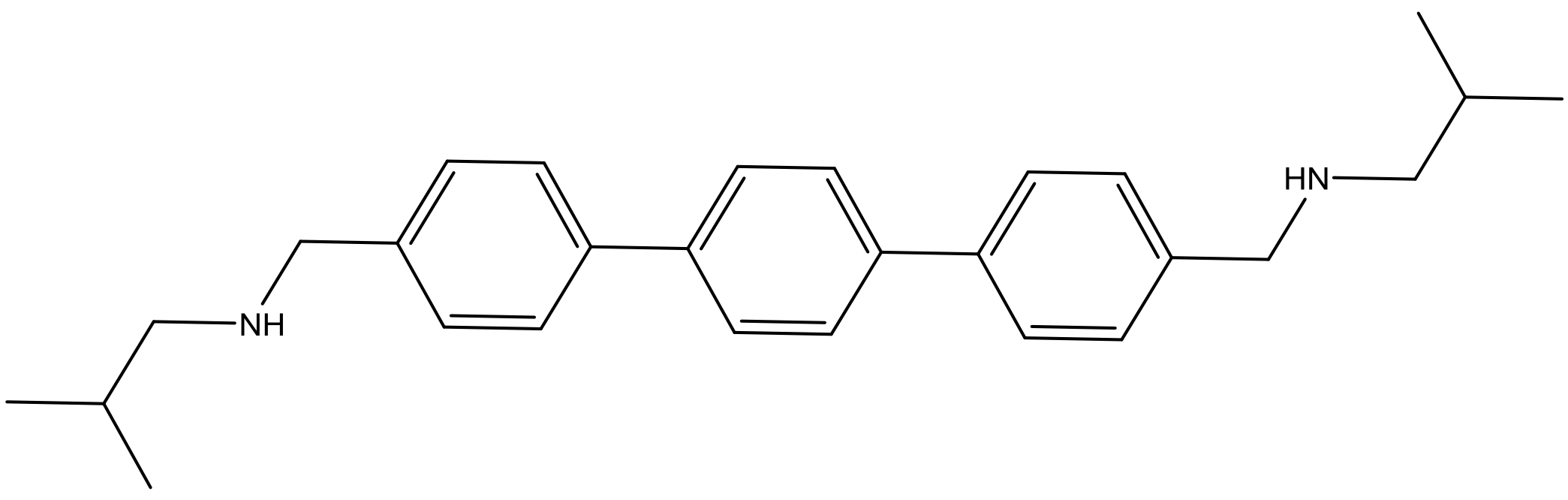

Isobutylamine (0.69 mL, 6.8 mmol) in methanol (5 mL) was added to a solution of terphenyl-4,4´-dicarbaldehyde (0.5 g, 1.7 mmol) in methanol (30 mL), and the reaction mixture was refluxed for 3 h under nitrogen, then allowed to stir at room temperature overnight. $NaBH_4$ (10 equivalents) was added and the reaction mixture was stirred for 12 h under nitrogen atmosphere. The reaction was quenched with water and the solvent was evaporated. The solid was extracted with chloroform ( 50 mL), washed with water and dried over anhydrous magnesium sulphate and evaporated. A light yellow liquid was obtained in 80 % yield (0.55 g). ESI-MS (sample dissolved in MeOH, run in MeOH): m/z = 401 $[M+H]^+$. $^1H$ NMR (500 MHz, 293K, $CDCl_3$): $\delta$= 1.3 (s, 12H), 1.4 (s br, 2H, NH), 2.61-2.72 (m, 6H), 3.67 (s, 4H), 7.53 (d, 4H) 7.60 (d, 4H), 7.72 (s, 4H).

**Synthesis of thread T3: $NC_5H_4$-$C_6H_4$-$CH_2$–NH-$CH_2CH_2$-Ph**

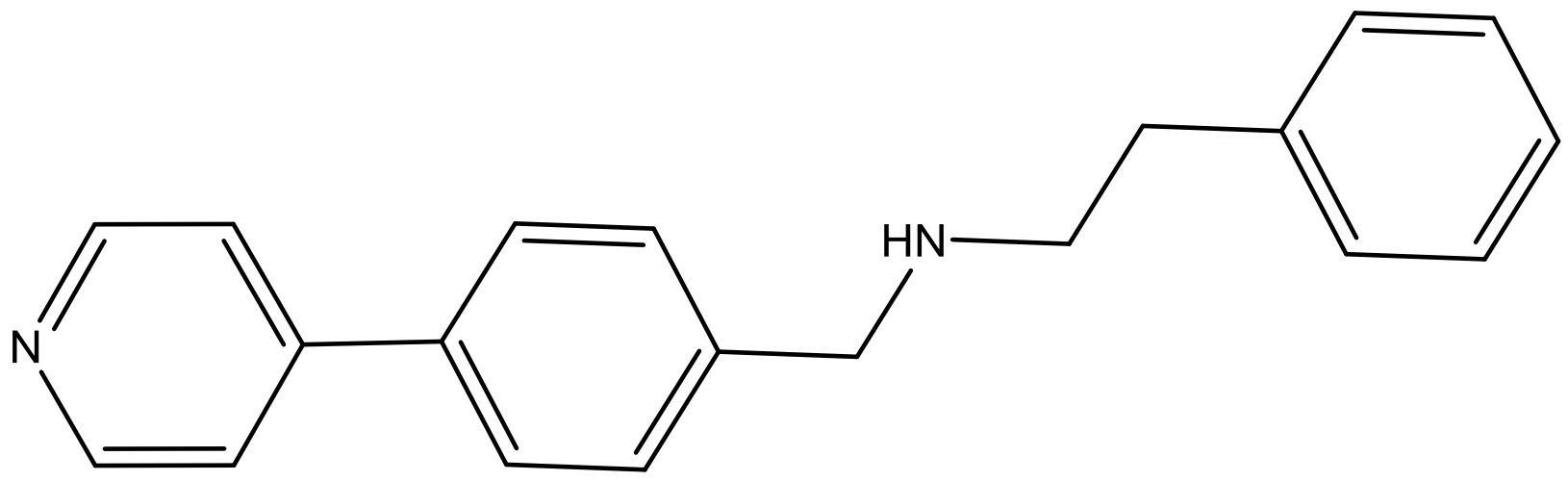

Phenethylamine (0.72 mL, 4.3 mmol) in methanol (5 mL) was added to a solution of 4-pyridine phenylaldehyde (0.84 g, 4.3 mmol) in methanol (30 mL), and the reaction mixture was refluxed for 3 h under nitrogen, then allowed to stir at room temperature overnight. $NaBH_4$ (5 equivalents) was added and reaction mixture was stirred during 12 h under nitrogen atmosphere. The reaction was quenched with water and the solvent was evaporated. The solid was extracted with chloroform (50 mL), washed with water and dried over anhydrous magnesium sulphate and evaporated. A light yellow liquid was obtained in 80 % yield (1.4 g). ESI-MS (sample dissolved in MeOH, run in MeOH): m/z = 289 $[M+H]^+$. $^1H$ NMR (400 MHz, 293K, $CDCl_3$): $\delta$ = 1.4 (s br, 2H, NH), 2.83-2.88 (t, 4H), 2.92-2.95 (t, 2H), 3.89 (s, 2H), 7.2-7.3 (m, 5H), 7.5 (d, 2H), 7.6 (d 2H) 7.7 (d, 2H), 8.7 (d, 2H).

**Synthesis of [3]rotaxane 1A: {$[H_2T1][Cr_7NiF_8(O_2C^tBu)_{16}]_2$} (1A)**

{[Ph-$CH_2$ $CH_2$-$NH_2$ -$CH_2$-$(C_6H_4)_3$-$CH_2$-$NH_2$ -$CH_2CH_2$-Ph]$[Cr_7NiF_8(O_2C^tBu)_{16}]_2$}**(1A)**

Pivalic acid (20.0 g, 195 mmol), **T1** (0.45 g, 0.91 mmol) and $CrF_3{\cdot}4H_2O$ (2.29 g, 12.6 mmol) were heated at 140°C with stirring in a Teflon flask for 0.5 h, then $[Ni_2(H_2O)(O_2C^tBu)_4(HO_2C^tBu)_4]$ 1.29 g, 1.36 mmol) was added. After 1 h the temperature of the reaction was increased to 160 °C for 20 h. The flask was cooled to room temperature, and then MeCN (35 mL) was added while stirring. The green microcrystalline product was collected by filtration, washed with a large quantity of MeCN, dried in air, and then extracted with hot toluene. Flash chromatography (toluene and then ethyl acetate/toluene (1/9)) afforded the desired [3]rotaxane as a green crystalline solid (1.5 g) in 34% (based on Cr) yield. Elemental analysis (%) calcd for $C_{196}H_{326}Cr_{14}F_{16}N_2Ni_2O_{64}$: Cr 14.90, Ni 2.40, C 48.20, H 6.73, N 0.57; found: Cr 14.93, Ni 1.89, C 47.70, H 6.81, N 0.50. X-ray quality crystals were obtained for **1A** by recrystallization from THF/MeCN.

**Synthesis of [3]rotaxane 1B: {$[H_2T2][Cr_7NiF_8(O_2C^tBu)_{16}]_2$} (1B)**

{$[(CH_3)_2CH\text{-}CH_2\text{-}NH_2\text{-}CH_2\text{-}(C_6H_4)_3\text{-}CH_2\text{-}NH_2\text{-}CH_2\text{-}CH(CH_3)_2][Cr_7NiF_8(O_2C^tBu)_{16}]_2$} **(1B)**
Pivalic acid (20.0 g, 195 mmol), **T2** (0.45 g, 0.91 mmol), and $CrF_3{\cdot}4H_2O$ (2.29 g, 12.6 mmol) were heated at 140 °C with stirring in a Teflon flask for 0.5 h, then $[Ni_2(H_2O)(O_2C^tBu)_4(HO_2C^tBu)_4]$ 1.29 g, 1.36 mmol) was added. After 1 h the temperature of the reaction was increased to 160 °C for 20 h. The flask was cooled to room temperature, and then acetonitrile (35 mL) was added while stirring. The green microcrystalline product was collected by filtration, washed with a large quantity of acetonitrile, dried in air, and then extracted with toluene. Flash chromatography (toluene and then ethyl acetate/toluene (1/9)) afforded the desired [3]-rotaxane as a green crystalline solid (1.6 g) in 37% (based on Cr) yield. Elemental analysis (%) calcd for $C_{188}H_{326}Cr_{14}F_{16}N_2Ni_2O_{64}$: Cr 15.20, Ni 2.45, C 47.16, H 6.86, N 0.59; found: Cr 14.48, Ni 2.01, C 47.64, H 7.03, N 0.50. X-ray quality crystals were obtained for **1B** by recrystallization from THF/MeCN.

**Synthesis of [3]rotaxane 1B-d** *(with deuterated pivalate ligands)*:

**{$[H_2T2][Cr_7NiF_8(O_2CC(CD_3)_3)_{16}]_2$} (1B-d)**

{$[(CH_3)_2CH\text{-}CH_2\text{-}NH_2\text{-}CH_2\text{-}(C_6H_4)_3\text{-}CH_2\text{-}NH_2\text{-}CH_2\text{-}CH(CH_3)_2][Cr_7NiF_8(O_2CC(CD_3)_3)_{16}]_2$}
**(1B-d)**
Deuterated pivalic acid $(CD_3)_3CCOOH$) (20.0 g, 195 mmol), **T2** (0.45 g, 0.91 mmol), and $CrF_3{\cdot}4H_2O$ (2.29 g, 12.6 mmol) were heated at 140 °C with stirring in a Teflon flask for 0.5 h, then $2NiCO_3{\cdot}3Ni(OH)_2\ 4H_2O$ (0.29 g, 0.49 mmol) was added. After 1 h the temperature of the reaction was increased to 160 °C for 20 h. The flask was cooled to room temperature, and then MeCN (35 mL) was added while stirring. The green microcrystalline product was collected by filtration, washed with a large quantity of MeCN, dried in air, and then extracted with toluene. Flash chromatography (toluene and then ethyl acetate/toluene (1/9)) afforded the desired [3]rotaxane as a green crystalline solid (1.4 g) in 30.5 % yield (based on Cr). Elemental analysis (%) calcd for $C_{188}H_{38}N_2Cr_{14}Ni_2F_{16}O_{64}D_{288}$: Cr 14.33, Ni 2.31, C 44.17, N 0.55; found: Cr 14.12, Ni 1.80, C 44.70, N 0.40.

**Synthesis of [3]rotaxane 1C: {$[H_2T1][Cr_7NiF_8(O_2CAd)_{16}]_2$} (1C)**

{[Ph-$CH_2CH_2$-NH-$CH_2$-$(C_6H_4)_3$-$CH_2$-NH-$CH_2CH_2$-Ph][$Cr_7NiF_8(O_2C_{11}H_{15})_{16}]_2$}**1C**

1-Adamantanecarboxylic acid ($AdCO_2H$) (10.0 g, 55.5 mmol), thread **T1** (0.4 g, 0.805 mmol), $CrF_3 \cdot 4H_2O$ (2.5 g, 13.81 mmol), $2NiCO_3 \cdot 3Ni(OH)_2$ $4H_2O$ (0.5 g, 0.851mmol) and 1,2-dichlorobenzene (6ml) were stirred together at 120 °C for 1h in open Erlenmeyer Teflon flask, then temperature was increased to 140 °C, after 1h the flask was connected to a slow flow of nitrogen and reaction continued for 24 h at 160 °C. The flask was cooled to room temperature and acetone (30ml) was added while stirring. The obtained product was filtered, washed with acetone, and then an extraction was performed using toluene (200ml) while stirring at 100 °C for 1h, the resulting solution was filtered, and filtrate concentrated under reduced pressure to *ca* 20 mL. This produced a precipitate, which was collected by filtration, washed with acetone and re-dissolved in hot toluene. Concentration of this solution by slow evaporation at R.T. produced after 2 weeks a crystalline product, including crystals suitable for X-ray study. Yield: 1.25 g (17%, based on Cr). Elemental analysis calculated (%) for $C_{388}H_{518}Cr_{14}F_{16}N_2Ni_2O_{64}$: Cr 9.86, Ni 1.59, C 63.12, H 7.07, N 0.38; found: Cr 9.88, Ni 1.62, C 63.04, H 7.10, N 0.39.

**Synthesis of [2]rotaxane {$[HT3][Cr_7NiF_8(O_2C^tBu)_{16}]$}**

[$NC_5H_4$-$C_6H_4$-$CH_2$–NH-$CH_2CH_2$-Ph][$Cr_7NiF_8(O_2C^tBu)_{16}$]}

Pivalic acid (20.0 g, 195 mmol), thread **T3** (0.5 g, 2.4 mmol), and $CrF_3 \cdot 4H_2O$ (2.2 g, 12 mmol) were heated at 140 °C with stirring in a Teflon flask for 0.5 h, then $[Ni_2(H_2O)(O_2C^tBu)_4(HO_2\ C^tBu)_4]$ (0.98 g, 1.03 mmol) was added. After 1 h the temperature of the reaction was increased to 160 °C for 24 h and the reaction was carried out under $N_2$ atmosphere. The flask was cooled to room temperature, and then MeCN (35 mL) was added while stirring. The green microcrystalline product was collected by filtration, washed with a large quantity of MeCN, dried in air, and then extracted with toluene. Flash chromatography (toluene and then ethyl acetate/toluene (3/7)) afforded the desired [2]-rotaxane as a green crystalline solid (0.6 g) in 14% yield (based on Cr). Elemental analysis (%) calcd for

$C_{100}H_{165}Cr_7F_8N_2NiO_{32}$: Cr 14.66, Ni 2.36, C 48.39, H 6.70, N 1.13; found: Cr 14.43, Ni 2.31, C 48.61, H 6.62, N 1.14. ESI-MS (sample dissolved in THF, run in MeOH): m/z = 2482 $[M+H]^+$; 2504 $[M+Na]^+$.

**Synthesis of [3]rotaxane 1D: $\{[HT3][Cr_7NiF_8(O_2C^tBu)_{16}]\}_2[Rh_2(O_2CMe)_4]$ (1D)**

$\{[NC_5H_4\text{-}C_6H_4\text{-}CH_2\text{–}NH\text{-}CH_2CH_2\text{-}Ph\ ][Cr_7NiF_8(O_2C^tBu)_{16}]\}_2[Rh_2(O_2CMe)_4]$ **1D** $\{[HT3][Cr_7NiF_8(O_2C^tBu)_{16}]\}$ (0.115 g, 0.046 mmol) and $[Rh_2(O_2CMe)_4]$ (0.01 g, 0.0226 mmol) were stirred in THF (10 mL) at R.T. for 24h. Toluene (10 ml) was added and the solvents removed while stirring in a flow of $N_2$ by evaporation. The residue was stirred with acetone (40 mL) under reflux for 5 min and the resulting solution filtered hot, then left at RT in a sealed flask. After 1h crystals started to form. After 2 weeks crystalline product (including suitable for X-ray structure study), and was collected by filtration, washed with acetone, and dried in *vacuo*. Yield 0.073 g (60% based on $[Rh_2(O_2CMe)_4]$). Elemental anal. calculated (%) for $C_{208}H_{342}Cr_{14}F_{16}N_4Ni_2O_{72}Rh_2$: Cr 13.47, Ni 2.17, Rh 3.81, C 46.21, H 6.38, N 1.04; Found: Cr 13.44, Ni 2.08, Rh 3.82, C 46.50, H 6.32, N 0.97

**Synthesis of dimer 2A: $\{[Cr_7NiF_3(Etglu)(O_2C^tBu)_{15}]_2(L1)\}$ (2A)**

Clathrochelate-based bipyridyl ligand L1 (0.060 g, 0.092 mmol) was added to a solution of $[Cr_7NiF_3(Etglu)(O_2C^tBu)_{15}(H_2O)]$ **2** (0.5 g, 0.225 mmol) in chloroform (15 mL), and the solution was stirred at RT overnight in sealed flask. The solvent was removed in a flow of $N_2$ and residue was stirred with acetone (20 mL) at RT for 24 h. Yellow-brown product was collected by filtration, washed with acetone (3 x 10 mL), then dissolved in pentane absolute (ca. 15 mL), filtered, and the filtrate diluted with acetone (ca. 15mL), then left for slow evaporation at RT in a flow of $N_2$. Crystals formed over two days (including suitable for X-ray structure study), and were collected by filtration, washed with acetone, and dried in *vacuo*. Yield: 0.41 g (88% based on the ligand L1); elemental analysis calcd (%) for ($C_{194}H_{330}B_2Cr_{14}F_6Fe_1N_{10}Ni_2O_{76}$): Cr 14.40, Ni 2.32, Fe 1.10 C 46.09, H 6.58, N 2.77; found: Cr 14.31, Ni 2.28, Fe 1.05, C 45.75, H 6.30, N 2.65.

**Synthesis of dimer 2B: $\{[Cr_7NiF_3(Etglu)(O_2C^tBu)_{15}]_2L2\}$ (2B)**

Clathrochelate-based bipyridyl ligand L2 (0.060 g, 0.061 mmol) and compound 2 $[Cr_7NiF_3(Etglu)(O_2C^tBu)_{15}(H_2O)]$ (0.35 g, 0.158mmol) were stirred in chloroform (18mL), at RT overnight in sealed flask. Then the solvent was removed in a flow of $N_2$ and residue was stirred with acetone absolute (10 ml) at RT for 24 h. Yellow-brown product was collected by filtration, washed with acetone (5 x 10 mL), then dissolved in pentane absolute (ca. 40 mL), filtered, and the filtrate diluted with acetone (ca. 20 mL), then left for slow evaporation at RT in a flow of $N_2$ Crystals formed over two days (including suitable for X-ray structure study), and were collected by filtration, washed with acetone, and dried in *vacuo*. Yield: 0.28 g (85% based on the ligand L2); elemental analysis calcd (%) for ($C_{218}H_{374}B_2Cr_{14}F_6Fe_1N_{10}Ni_2O_{76}$): Cr 13.51, Ni 2.18, Fe 1.04 C 48.59, H 7.00, N 2.60; found: Cr 13.64, Ni 2.15, Fe 1.02, C 48.20, H 7.00, N 2.49

**Synthesis of dimer 2C: {[$Cr_7NiF_3$(Etglu)($O_2C^tBu$)$_{15}$]$_2$L3} (2C)**

The same procedure as for 2B with the exception that clathro-chelate L3 was used instead of L2. Crystals formed over one week (including suitable for X-ray structure study), and were collected by filtration, washed with acetone, and dried in *vacuo*. Yield: 0.11 g (40% based on the ligand C); elemental analysis calcd (%) for($C_{218}H_{358}B_4Cr_{14}F_6Fe_2N_{16}Ni_2O_{82}$): Cr 12.93, Ni 2.09, Fe 1.98 C 46.51, H 6.41, N 3.98; found: Cr 13.04, Ni 2.05, Fe 2.30, C 46.20, H 6.58, N 3.83

## 2 Single Crystal *X*-ray Crystallography

Data for **1D** were collected using synchrotron radiation (λ = 0.8266 Å) on a Bruker Apex II diffractometer at the Advanced Light Source. Data for **1A**, **1B** and **2A** were collected on a Bruker X8 Prospector 3-circle diffractometer with a copper microfocus source and an APEX II CCD detector. Data for **1C** and **2B** were collected at 100 K on beamline I19 at Diamond Light Source (H. Nowell, S. A. Barnett, K. E. Christensen, S. J. Teat and D. R. Allan, *J. Synch. Rad.*,2012, **19**, 435-441) using a Rigaku Saturn Kappa diffractometer and the data reduced using Agilent Technologies CrysAlisPro. Data for **2C** were collected using synchrotron radiation (λ = 0.7749 Å) on a Bruker Apex II diffractometer. The structures were solved and refined using SHELX97 and SHELX-2013 (G. M. Sheldrick , *Acta Cryst.*, 2008, **A64**, 112-122 and O. V. Dolomanov, L. J. Bourhis, R. J. Gildea, J. A. K. Howard and H. Puschmann , *J. Appl. Cryst.*, 2009, **42**, 339-341).

Crystal data and refinement parameters are given in Table S1. CCDC 1039429-1039435 contain the supplementary crystallographic data for this paper. These data can be obtained free of charge from The Cambridge Crystallographic Data Centre via www.ccdc.cam.ac.uk/data_request/cif

***Table S1:*** Crystallographic data for **1A, 1B, 1C, 1D, 2A, 2B** and **2C**

| | **1A** **AF297** | **1B** **AF377** | **1C** **GT76_14** | **1D** **gt-259-13** |
|---|---|---|---|---|
| empirical formula | $C_{196}H_{323}Cr_{14}F_{16}N_2Ni_2O_{64}$ | $C_{196}H_{342}Cr_{14}F_{16}N_2Ni_2O_{64}$ | $C_{416}H_{550}Cr_{14}F_{16}N_2Ni_2O_{64}$ | $C_{220}H_{350}Cr_{14}F_{16}N_4Ni_4O_{72}Rh_2$ |
| formula weight | 4880.96 | 4947.99 | 7751.96 | 5558.26 |
| temperature / K | 100(2) | 150(2) | 100(2) | 100(2) |
| crystal system | Triclinic | monoclinic | triclinic | orthorhombic |
| space group | *P*-1 | *C*2/*c* | *P*-1 | *Pbcn* |
| *a* / Å | 16.4604(7) | 56.6600(18) | 19.4348(10) | 79.046(3) |
| *b* / Å | 29.4960(11) | 16.6410(4) | 20.605(2) | 29.2781(12) |
| *c* / Å | 31.4568(14) | 31.0446(8) | 30.172(2) | 28.0617(11) |
| α / ° | 117.088(4) | 90 | 76.824(7) | 90 |
| β / ° | 93.268(4) | 110.0710(10) | 83.141(5) | 90 |
| γ / ° | 98.284(3) | 90 | 63.719(7) | 90 |
| volume / $Å^3$ | 13324.6(11) | 27493.6(13) | 10546.1(15) | 64944(4) |
| *Z* | 2 | 4 | 1 | 8 |
| *λ*/Å | 1.54178 | 1.54178 | 0.6889 | 0.8266 |
| $\rho_{calc}$ / $mg.mm^{-3}$ | 1.217 | 1.195 | 1.221 | 1.158 |
| μ / $mm^{-1}$ | 5.303 | 5.158 | 0.468 | 1.098 |
| reflections collected | 39628 | 56903 | 72108 | 262506 |
| independent reflections | 24554 [R(int) = 0.1696] | 19622 [R(int) = 0.097] | 30005 [R(int) = 0.0685] | 39770 [R(int) = 0.0817] |
| data / restraints / parameters | 24554 / 11496 / 2587 | 19622 / 2045 /1328 | 30005 / 5876 / 2444 | 39770 /9748 / 2920 |
| goodness-of-fit on $F^2$ | 1.034 | 1.068 | 1.312 | 1.083 |
| final R indexes [I≥2σ(I)] | $R_1$ = 0.170, $wR_2$ = 0.390 | $R_1$ = 0.1564, $wR_2$ = 0.388 | $R_1$ = 0.136, $wR_2$ = 0.3666 | $R_1$ = 0.135, $wR_2$ = 0.358 |
| final R indexes [all data] | $R_1$ = 0.257, $wR_2$ = 0.461 | $R_1$ = 0.1797, $wR_2$ = 0.4036 | $R_1$ = 0.1857, $wR_2$ = 0.4149 | $R_1$ = 0.1444, $wR_2$ = 0.3644 |
| largest diff. peak / hole /$e.A^{-3}$ | 1.48 / -1.225 | 1.96 / -1.33 | 1.962 / -1.649 | 1.168 / -1.263 |

| | **2A**<br>Xrepw134-GT-202-12 | **2B**<br>GT36-13 | **2C**<br>GT-39-13a |
|---|---|---|---|
| empirical formula | $C_{194}H_{330}B_2Cr_{14}F_6FeN_{10}Ni_2O_{76}$ | $C_{218}H_{368}B_2Cr_{14}F_6FeN_{10}Ni_2O_{76}$ | $C_{221}H_{363}B_4Cr_{14}F_6Fe_2N_{17}Ni_2O_{82}$ |
| formula weight | 5055.46 | 5382.09 | 5755.52 |
| temperature / K | 150(2) | 100(2) | 100(2) |
| crystal system | orthorhombic | orthorhombic | orthorhombic |
| space group | $P2_12_12_1$ | $P2_12_12$ | $P2_12_12$ |
| *a* / Å | 23.1181(8) | 28.2831(9) | 29.1790(8) |
| *b* / Å | 25.9891(11) | 70.444(4) | 72.351(2) |
| *c* / Å | 47.1671(18) | 16.6428(5) | 16.7689(5) |
| α / ° | 90 | 90 | 90 |
| β / ° | 90 | 90 | 90 |
| γ / ° | 90 | 90 | 90 |
| volume / $Å^3$ | 28338.9(19) | 34331(2) | 35401.2(17) |
| *Z* | 4 | 4 | 4 |
| λ/Å | 1.54178 | 0.6889 | 0.7749 |
| $\rho_{calc}$ / $mg.mm^{-3}$ | 1.185 | 1.041 | 108 |
| μ / $mm^{-1}$ | 5.382 | 0.547 | 0.848 |
| reflections collected | 78812 | 181314 | 200003 |
| independent reflections | 50233<br>[R(int) = 0.1051] | 41702<br>[R(int) = 0.1379] | 43095<br>[R(int) = 0.0598] |
| data / restraints / parameters | 50233 / 1318 / 2746 | 41702 / 3878 / 2547 | 43095 /4313 / 3116 |
| goodness-of-fit on $F^2$ | 1.018 | 1.264 | 1.131 |
| final R indexes [I≥2σ(I)] | $R_1$ = 0.0904,<br>$wR_2$ = 0.2139 | $R_1$ = 0.1514,<br>$wR_2$ = 0.3732 | $R_1$ = 0.0909,<br>$wR_2$ = 0.2532 |
| final R indexes [all data] | $R_1$ = 0.1492,<br>$wR_2$ = 0.2536 | $R_1$ = 0.2121,<br>$wR_2$ = 0.4218 | $R_1$ = 0.0934,<br>$wR_2$ = 0.2554 |
| largest diff. peak / hole /$e.A^{-3}$ | 2.54 / -0.62 | 1.10 / -1.17 | 1.723 / -0.785 |

# 3 ESR spectroscopy

## 3.1 Field sweep spectroscopy

Continuous wave ESR spectra and echo-detected magnetic field sweeps were recorded at X-band (9.5 GHz). For each compound, spectra were fitted with a simultaneous set of Hamiltonian parameters to obtain an accurate description of the system. The unresolved hyperfine couplings were included in the simulation as an *H*-strain parameter.

The simulation parameters are tabulated in Table S2. X-band CW data and simulations for **1B**, **1C**, **1D**, **2B** and **2C** are plotted in Figure S1 (see main manuscript for **1A** and **2A**).

***Table S2:*** *Spin Hamiltonian parameters derived from magnetic field sweep ESR*

| Compound | *g*-tensor | *g*-strain | *H*-strain |
|---|---|---|---|
| **1A** | [1.788 1.788 1.748] | [0.02 0.02 0.06] | [250 250 250] |
| **1B** | [1.788 1.788 1.748] | [0.02 0.02 0.06] | [250 250 250] |
| **1C** | [1.788 1.788 1.748] | [0.02 0.02 0.06] | [250 250 250] |
| **1D** | [1.788 1.788 1.748] | [0.02 0.02 0.06] | [250 250 250] |
| **2A** | [1.83 1.83 1.76] | [0.02 0.02 0.03] | [200 200 200] |
| **2B** | [1.83 1.83 1.77] | [0.02 0.02 0.03] | [200 200 200] |
| **2C** | [1.83 1.83 1.76] | [0.02 0.02 0.03] | [200 200 200] |

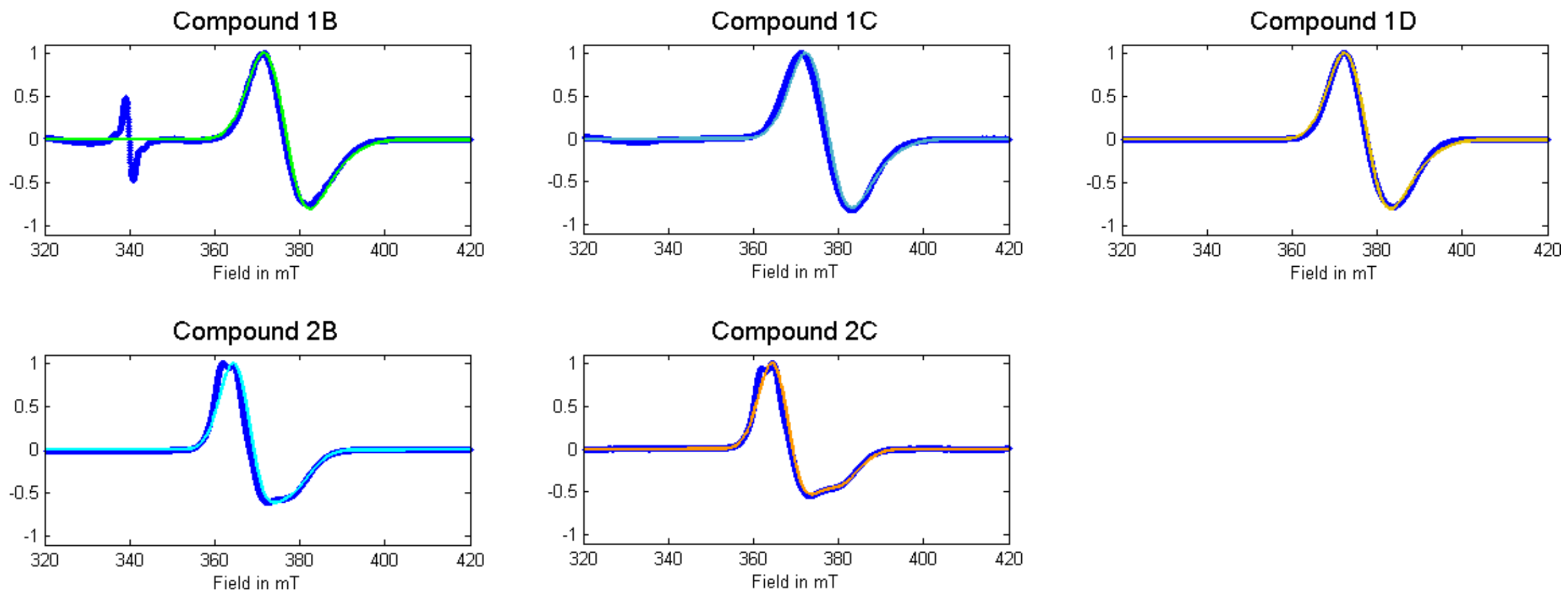

***Figure S1:*** *X-band continuous wave ESR and simulations*

### 3.2 DEER experiments

Double Electron – Electron Resonance (DEER) was measured for all compounds. All DEER experiments used $\nu_1$ detection pulses ($\pi/2$ and $\pi$) of 40 ns and $\nu_2$ pump pulses of length of 24 ns. Compounds **1A**, **1Bd**, **1C** and **2A** were measured using 4-pulse DEER. Owing to the shorter coherence times in samples **1D** and **2B** it was necessary to use a combination of 3- and 4- pulse DEER techniques. In these cases, a 4-pulse trace with a short acquisition window was recorded and carefully combined with a 3-pulse trace with a longer acquisition window using the DEER-stitch method described in Lovett et al., *Journal of Magnetic Resonance* **223** 98 (2012), to reconstruct the zero time on the longer 3-pulse trace. Sample **2C** was measured using 3-pulse DEER only.

Data were processed using DeerAnalysis (Jeschke et al., *Appl. Magn. Reson.* **30**, 473 (2006)). Using this package, the raw data were smoothed, filtered to remove the effects of electron spin echo envelope modulation (ESEEM) arising from protons, and background-corrected to account for inter-dimer interactions in the three-dimensional homogenous distribution. The sample concentrations were sufficiently small that the background correction was at all times small.

In each of the following figures, open circles represent raw data and solid lines represent smoothed filtered data. At low magnetic fields, rings with effective $g$-factors close to $g_\perp$ (i.e. dimers with inter-ring axes approximately perpendicular to the field) are excited; as the magnetic field increases, the ring orientations selected approach those with effective $g$-factors closer to $g_\parallel$ (i.e. dimers are selected whose axes approach the direction parallel to the field).

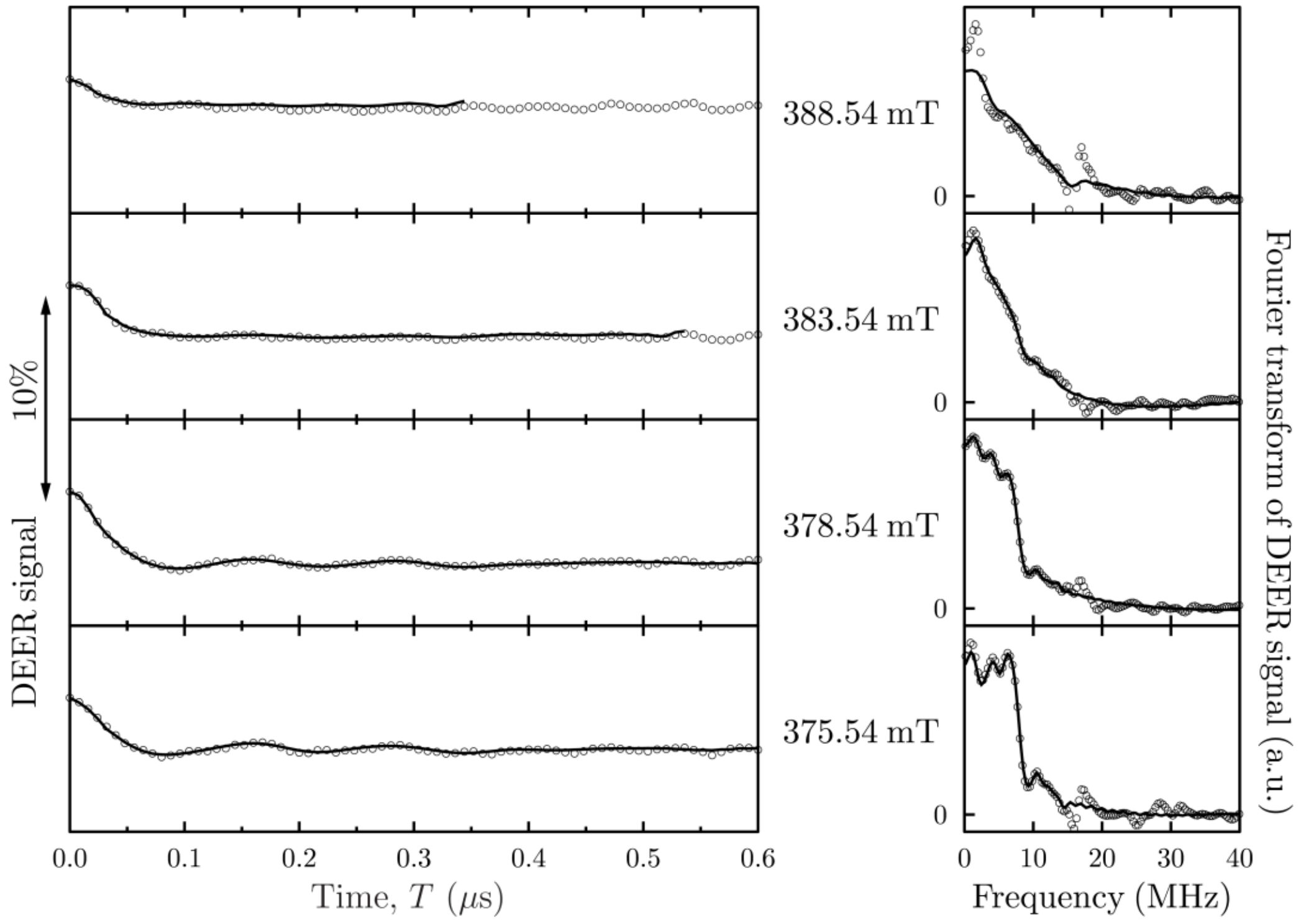

***Figure S2:*** *DEER data for compound **1B**. $\nu_1$ = 9.5287 GHz. $\nu_2 = \nu_1$ – 80MHz = 9.4487 GHz. Additional intensity at low frequencies may be associated with the deuterium content in this compound. Deuterium nuclei cause electron spin echo envelope modulations (ESEEM) occur with an oscillation period of about 456 ns at X-band frequencies, corresponding to a frequency of about 2.19 MHz.*

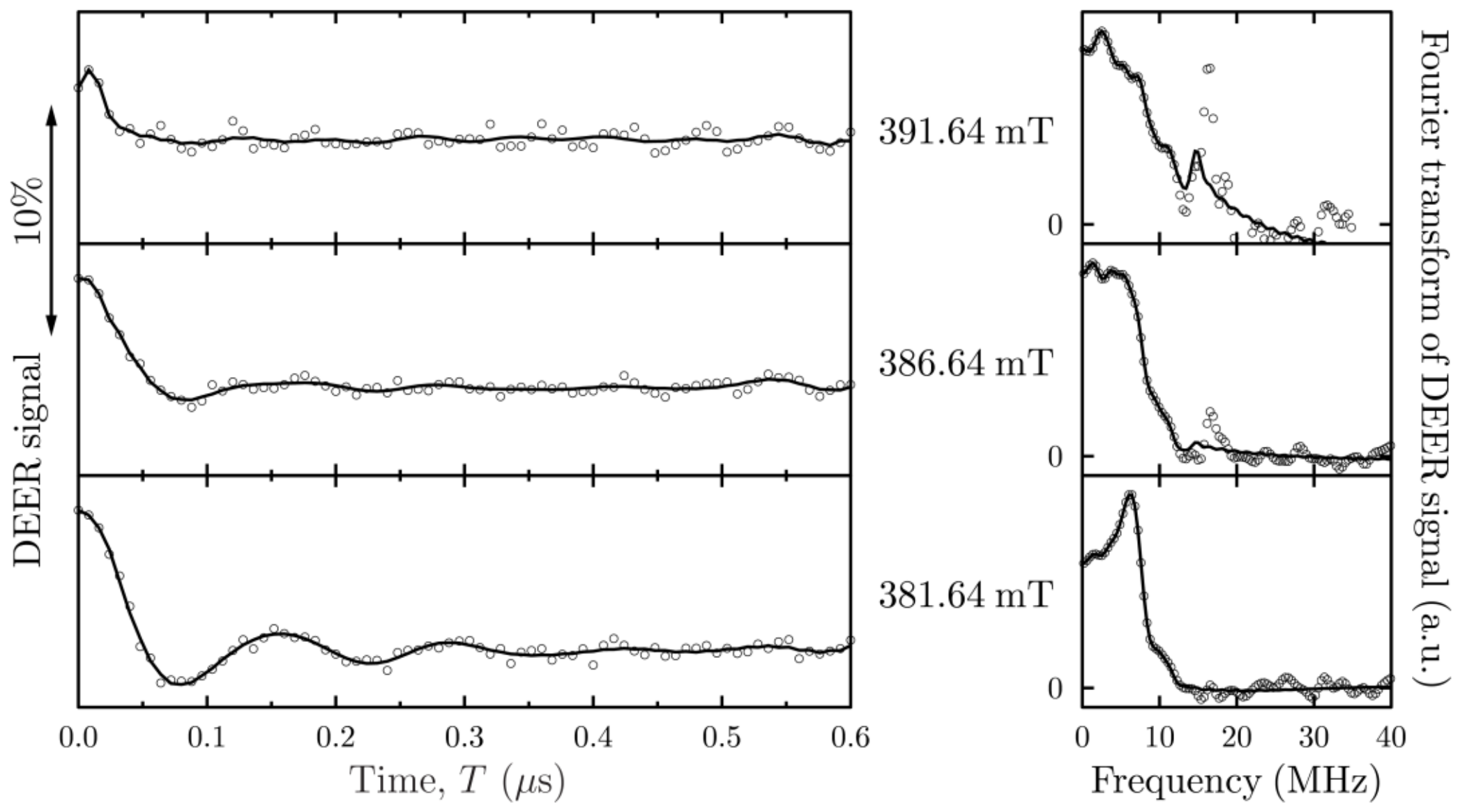

***Figure S3:*** *DEER data for compound **1C**. $\nu_1$ = 9.5577 GHz. $\nu_2 = \nu_1$ – 80MHz = 9.4777 GHz.*

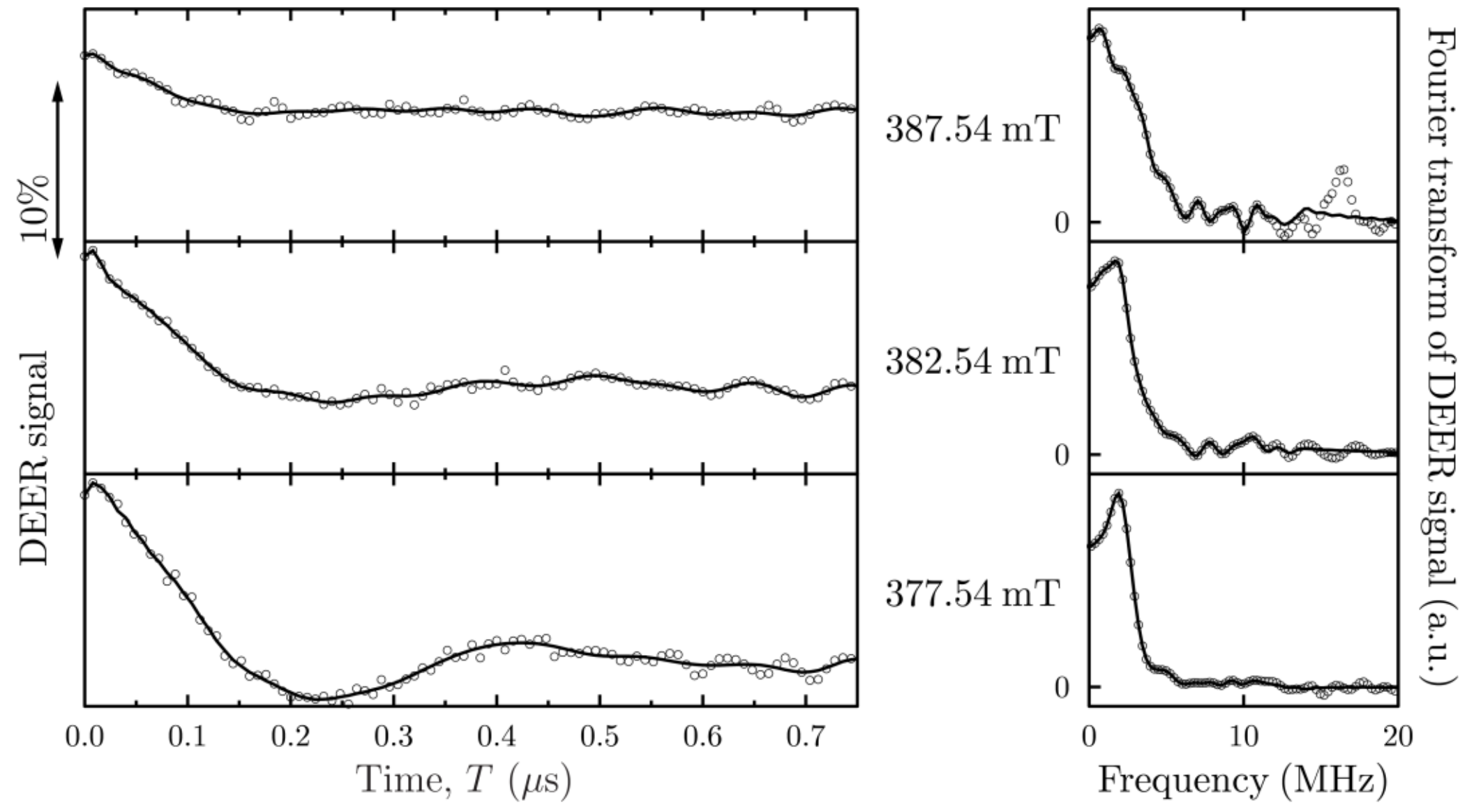

***Figure S4:*** *DEER data for compound* ***1D****.* $\nu_1$ *= 9.4888 GHz.* $\nu_2 = \nu_1$ *– 80MHz = 9.4088 GHz*

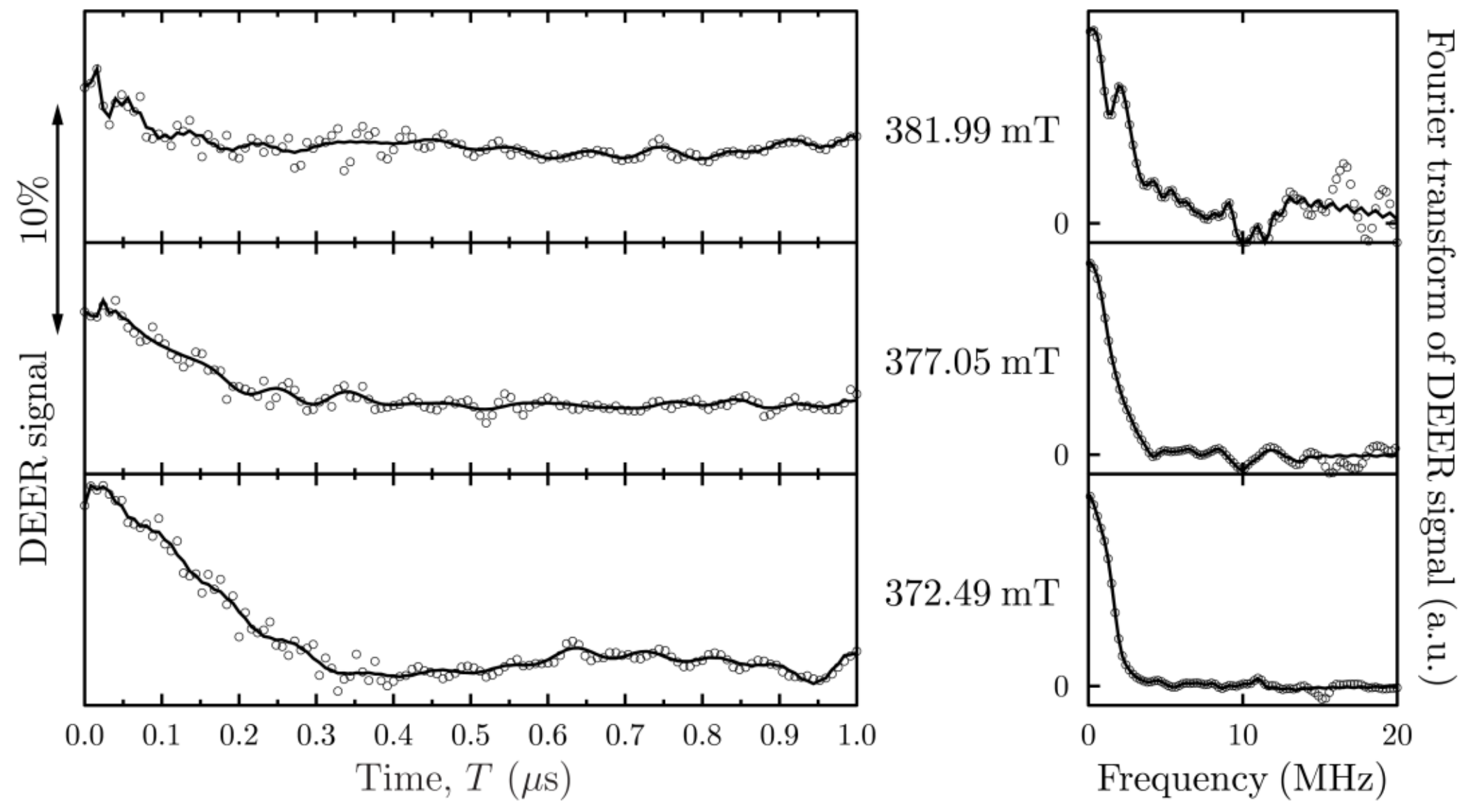

***Figure S5:*** *DEER data for compound* ***2B****.* $\nu_1$ *= 9.5100 GHz.* $\nu_2 = \nu_1$ *– 80MHz = 9.4300 GHz****.*** *Distortions close to T=0 arise from overlaps of the pump and probe pulses in time.*

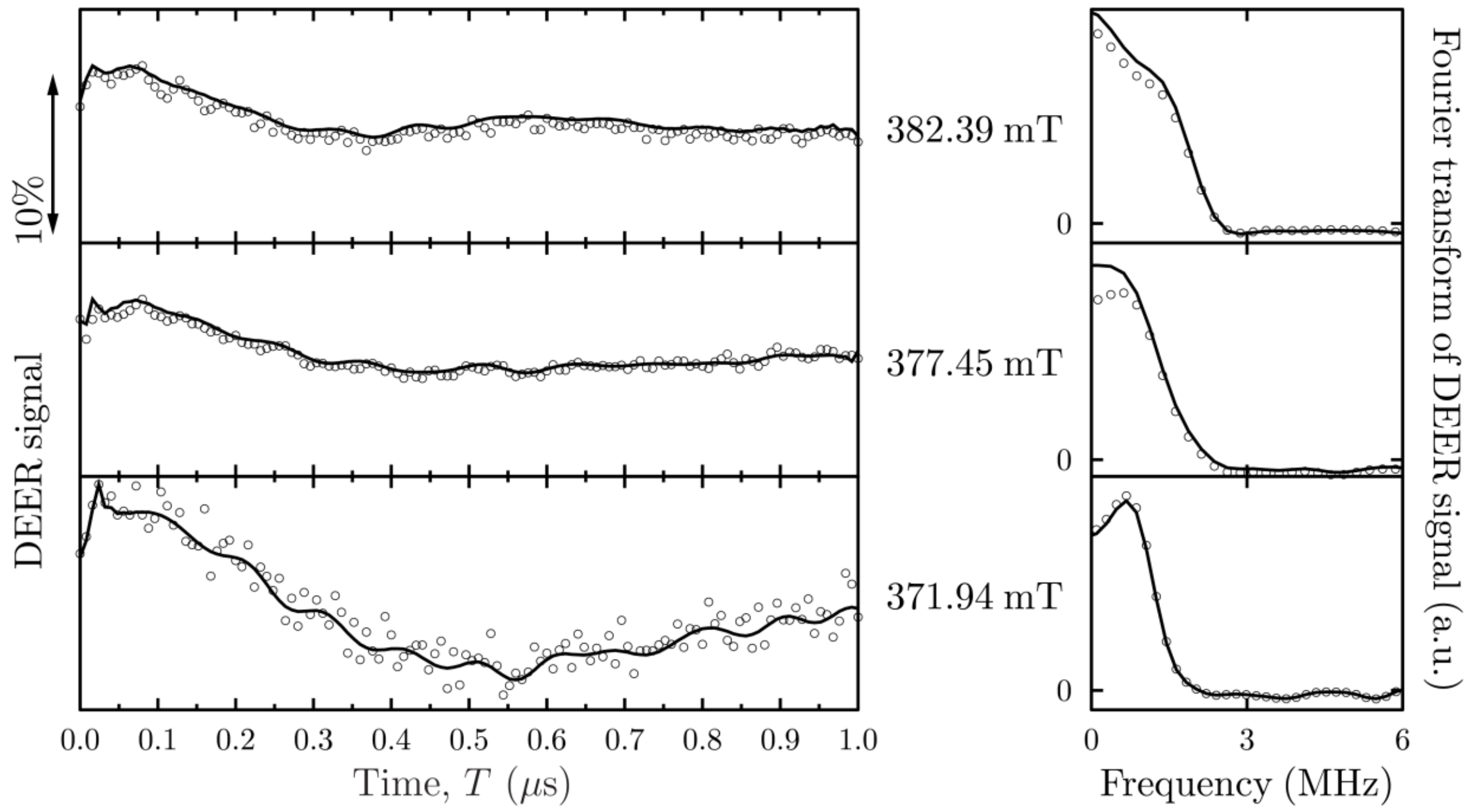

***Figure S6:*** *DEER data for compound* ***2C****.* $\nu_1$ *= 9.5200 GHz.* $\nu_2 = \nu_1$ *– 80MHz = 9.4400 GHz. Distortions close to T=0 arise from overlaps of the pump and probe pulses in time.*